\documentclass[preprint,3p,times,12pt]{elsarticle}  

 \pdfoutput=1

\usepackage{hyperref,color,amsmath,subcaption}                                                                                                                                                            
                                                                                 
\journal{Astroparticle Physics}                                                                       
         
\include{journalMacros}

\usepackage{xspace}
\def\Offline{\mbox{$\overline{\textrm%
      {Off}}$\hspace{.05em}\protect\raisebox{.4ex}%
    {$\protect\underline{\textrm{line}}$}}\xspace}%

\begin{document}                                                                                                   
                                                                                                                   
\begin{frontmatter}                                                                                                
                                                                                                                   
\title{The First Full-Scale Prototypes of the Fluorescence detector Array of Single-pixel Telescopes}                  
                                                                                                                   
\author[1]{M.~Malacari}                                                               
\ead{mmalacari@kicp.uchicago.edu}                                                          
\author[1]{J.~Farmer}
\ead{farmerjm@uchicago.edu}
\author[2]{T.~Fujii}

\author[3]{J.~Albury}
\author[3]{J.A.~Bellido}
\author[4]{L.~Chytka}
\author[4]{P.~Hamal}
\author[4]{P.~Horvath}
\author[4]{M.~Hrabovsk\'{y}}
\author[5]{D.~Mandat}
\author[6]{J.N.~Matthews}
\author[4]{L.~Nozka}
\author[4]{M.~Palatka}
\author[4]{M.~Pech}
\author[1]{P.~Privitera}
\author[4]{P.~Schov\'{a}nek}
\author[1]{R.~\v{S}m\'{i}da}
\author[6]{S.B.~Thomas}
\author[5]{P.~Travnicek}
            
\address[1]{Kavli Institute for Cosmological Physics, University of Chicago, Chicago, IL, USA}                                
\address[2]{Hakubi Center for Advanced Research and Graduate School of Science, Kyoto University, Sakyo, Kyoto, Japan}
\address[3]{Department of Physics, University of Adelaide, Adelaide, S.A., Australia}
\address[4]{Palack\'{y} University, Joint Laboratory of Optics of PU and CAS, Olomouc, Czech Republic} 
\address[5]{Institute of Physics of the Academy of Sciences of the Czech Republic, Prague, Czech Republic}
\address[6]{High Energy Astrophysics Institute, University of Utah, Salt Lake City, UT, USA}

                                                                                                                   
                                                                                                                   
\begin{abstract}   
The Fluorescence detector Array of Single-pixel Telescopes (FAST) is a design concept for a next-generation UHECR observatory, addressing the requirements for a large-area, low-cost detector suitable for measuring the properties of ultra-high energy cosmic rays (UHECRs), having energies exceeding $10^{19.5}$\,eV, with an unprecedented aperture. 
We have developed a full-scale prototype consisting of four 200\,mm diameter photo-multiplier tubes at the focus of a segmented mirror of 1.6\,m in diameter. 
In October 2016, September 2017, and September 2018 we installed three such prototypes at the Black Rock Mesa site of the Telescope Array experiment in central Utah, USA. All three telescopes have been steadily taking data since installation.
We report on the design and installation of these prototypes, and present some preliminary results, including measurements of artificial light sources, distant ultraviolet lasers, and UHECRs.
Furthermore, we discuss some additional uses for these simplified low-cost fluorescence telescopes, including the facilitation of a systematic comparison of the transparency of the atmosphere above the Telescope Array experiment and the Pierre Auger Observatory, a study of the systematic uncertainty associated with the existing fluorescence detectors of these two experiments, and a cross-calibration of their energy and $X_{\text{max}}$ scales.                                                       
\end{abstract}                                                                                                     
                                                                                                                   
                                                                                                                   
\end{frontmatter}

\section{Introduction}
	
While discoveries made over the past decade have transformed our understanding of ultra-high energy cosmic rays (UHECRs) and their sources, their nature and origin remain a mystery. Upon interaction with the earth's atmosphere, a UHECR produces a cascade of secondary particles known as an extensive air shower (EAS). By measuring the properties of these showers, we can determine the nature of the primary particles that produced them, such as their arrival direction, mass composition, and energy spectrum. Because of the exceptionally low  UHECR flux at the highest energies (0.5 per square kilometre per year above an energy of $10^{19.5}$\,eV), it is necessary to instrument a very large area in order to collect a suitably large dataset for meaningful statistical analysis. There is therefore a strong motivation to develop detectors that are low-cost, robust, and autonomous, which can be deployed in very large ground arrays to directly measure the energy and mass composition of the highest energy cosmic rays.

The largest present-day cosmic ray experiments are hybrid detectors that employ a combination of two techniques: surface detection (sampling the lateral distribution of secondary EAS particles at ground level) and fluorescence detection (observing the faint isotropically-emitted ultraviolet light produced during the de-excitation of atmospheric nitrogen). The Telescope Array experiment~\cite{bib:TA} (TA), spanning $700\,\text{km}^2$ in the desert of Utah, USA, and the Pierre Auger Observatory~\cite{bib:ThePierreAuger:2015rma} (Auger), spanning $3000\,\text{km}^2$ in the province of Mendoza, Argentina, both instrument a very large area with a grid of surface detector stations overlooked by a set of fluorescence telescopes.  

Surface detectors feature an exceptionally high duty cycle ($\sim100\%$) and excel in providing information about the lateral distribution of particles in the EAS at ground level~\cite{bib:augerSD,bib:tasd}.  Fluorescence detectors provide a calorimetric measurement of a shower's energy by collecting fluorescence light emitted during its longitudinal development. This method does not depend on extrapolation of accelerator-tuned hadronic interaction models to higher energies and is therefore a more accurate estimator of the shower energy. Observing a shower's development directly also provides another distinct advantage: observation of the depth of maximum development ($X_{\text{max}}$), a parameter indicative of the primary particle's mass.  In spite of these benefits, fluorescence detectors suffer from a significantly lower duty cycle $(\sim15\%)$ and reduced directional coverage~\cite{bib:augerFD,bib:tafd}. Using a subsample of showers detected in coincidence with both a surface and fluorescence detector, the calorimetric measurement of the shower energy provided by the fluorescence detector can be used to calibrate the energy scale of the surface detector using a suitable measured observable. This is known as hybrid detection, and is employed by both the Auger and TA collaborations to calibrate the energy measured by the high duty cycle surface array. 

Recent years have seen significant progress in the field, with advances in detector technology, reductions in systematic uncertainties, and increased exposures enabling the UHECR energy spectrum, composition, and anisotropies to be measured with increased resolution above $10^{17}$\,eV. Both TA and Auger have measured the energy spectrum up to $\sim10^{20}$\,eV, with clear indications of a break at around $10^{18.7}$\,eV (the ankle), and a flux suppression above $10^{19.7}$\,eV often attributed to energy loss through interactions with blue-shifted (in the centre of mass of the CR particle) cosmic microwave and infrared background photons through the GZK process~\cite{bib:gzk1,bib:gzk2}. Measurements of the elongation rate ($X_{\text{max}}$ as a function of energy) by both experiments indicate a predominantly light composition around the ankle, while above $10^{19}$\,eV Auger shows a decrease in the growth of $X_{\text{max}}$ with energy, as well as a decrease in RMS($X_{\text{max}}$), indicating a gradual increase in the average cosmic ray mass. Recent results have provided increasing motivation to probe the energy range above 100\,EeV, such as the Auger surface detector's hints at a lighter composition above $10^{19.5}$\,eV~\cite{Aab:2017cgk}. Below the ankle both TA and Auger measure arrival directions that are highly isotropic, with warm- and hot-spots appearing at higher energies due to the smearing of point sources by both Galactic and extragalactic magnetic fields. TA has recorded an excess above isotropic background expectations above 57\,EeV~\cite{bib:tahotspot}, while a blind search using a combination of both Auger and TA data shows an excess above this energy in a $20^{\circ}$ search window with a $2.2\sigma$ post-trial significance~\cite{bib:aniso_wg2018}. In 2017 Auger reported on a large-scale dipole above 8\,EeV with a $5.2\sigma$ significance, pointing $125^{\circ}$ away from the galactic centre, suggesting an extragalactic origin for the highest energy particles~\cite{bib:augerdipole, bib:augerdipole2}. To further advance and establish the field of charge particle astronomy, the next generation of ground-based UHECR detectors will require an unprecedented aperture, which is larger by an order of magnitude; mass composition sensitivity above 100\,EeV; and energy, $X_{\text{max}}$, and angular resolutions that are comparable to those of current-generation experiments.
 
The Fluorescence detector Array of Single-pixel Telescopes (FAST)~\cite{bib:firstfast} is an R\&D project aimed at developing a next-generation cosmic ray detector.  It is a low-cost fluorescence telescope sensitive to UHECRs with energies greater than $\sim 10^{19.5}$\,eV.  The main features of its design are a portable, compact mechanical structure and a camera consisting of four $200\,$\,mm diameter PMTs, in contrast to the expensive, highly-pixelated cameras used by both Auger and TA. This coarse granularity provides insufficient information for geometrical reconstruction with a single FAST telescope (in contrast to the directional timing information provided by pixels with small angular coverage), but when paired with geometry from an array of surface detectors, FAST could provide an independent measurement of the shower energy and $X_{\text{max}}$.  Further, if multiple FAST telescopes were deployed in an array over a large area, one could employ a top-down reconstruction method that simultaneously fits a shower's geometry and longitudinal development against a library of simulated templates. The FAST design may be an attractive option not only for future UHECR experiments, but also for upgrades to existing UHECR observatories. For example, it could be used at Auger to increase the number of showers detected in stereo with more than one fluorescence telescope.

A first test of the FAST concept was performed in 2014 using a single 200\,mm diameter PMT at the focus of the 1\,m$^2$ Fresnel lens system of the TA-EUSO optics at the TA site. With this first prototype, we detected 16 highly-significant UHECR shower signals and demonstrated excellent operational stability under conditions typical of field deployment~\cite{bib:firstfast}. Motivated by these encouraging results, we developed and installed three full-scale FAST prototypes at the Black Rock Mesa site of the Telescope Array experiment. In this paper, we report on the design and installation of these full-scale prototypes and present some preliminary results, including measurements of artificial light sources, distant ultraviolet lasers, and UHECRs. We present the FAST telescope optical, mechanical, and electrical design in Section~\ref{sec:design}. The three FAST prototypes installed at the TA site are described in Section~\ref{sec:TA_FAST}, along with details of their operation, ancillary instruments, and some preliminary recorded data. Section~\ref{sec:simRec} describes the FAST event simulation and reconstruction software, and a procedure to infer the vertical atmospheric transparency from measurements of a distant ultraviolet laser. The future of the FAST project is reported in Section~\ref{sec:future}, including the installation of an identical fourth telescope at the Auger site in the southern hemisphere. Finally, conclusions are drawn in Section~\ref{sec:conclusion}.

\section{The FAST prototype telescopes}
\label{sec:design}

\subsection{Telescope design}

A lensless Schmidt-type optical design was adopted for the full-size FAST prototype~\cite{bib:fastoptics}. In a typical Schmidt telescope a corrector plate is placed at the entrance aperture (located at the mirror's radius of curvature, a distance of $2f$, where $f$ is the focal length) to facilitate the control of off-axis aberrations: coma and astigmatism. The coarse granularity of the FAST camera, having only four PMTs each covering an angular field-of-view of $\sim15^{\circ}$, allows the requirements on the size and shape of the telescope's point spread function to be relaxed. The FAST prototype telescope therefore forgoes the use of a corrector plate, utilises a reduced-size mirror, and uses a shorter distance between the mirror and the camera relative to a regular Schmidt telescope, with the entrance aperture located closer to the focal surface.

\begin{figure}[t]
    \begin{subfigure}[t]{0.5\textwidth}
        \centering
        \includegraphics[width=1.\linewidth]{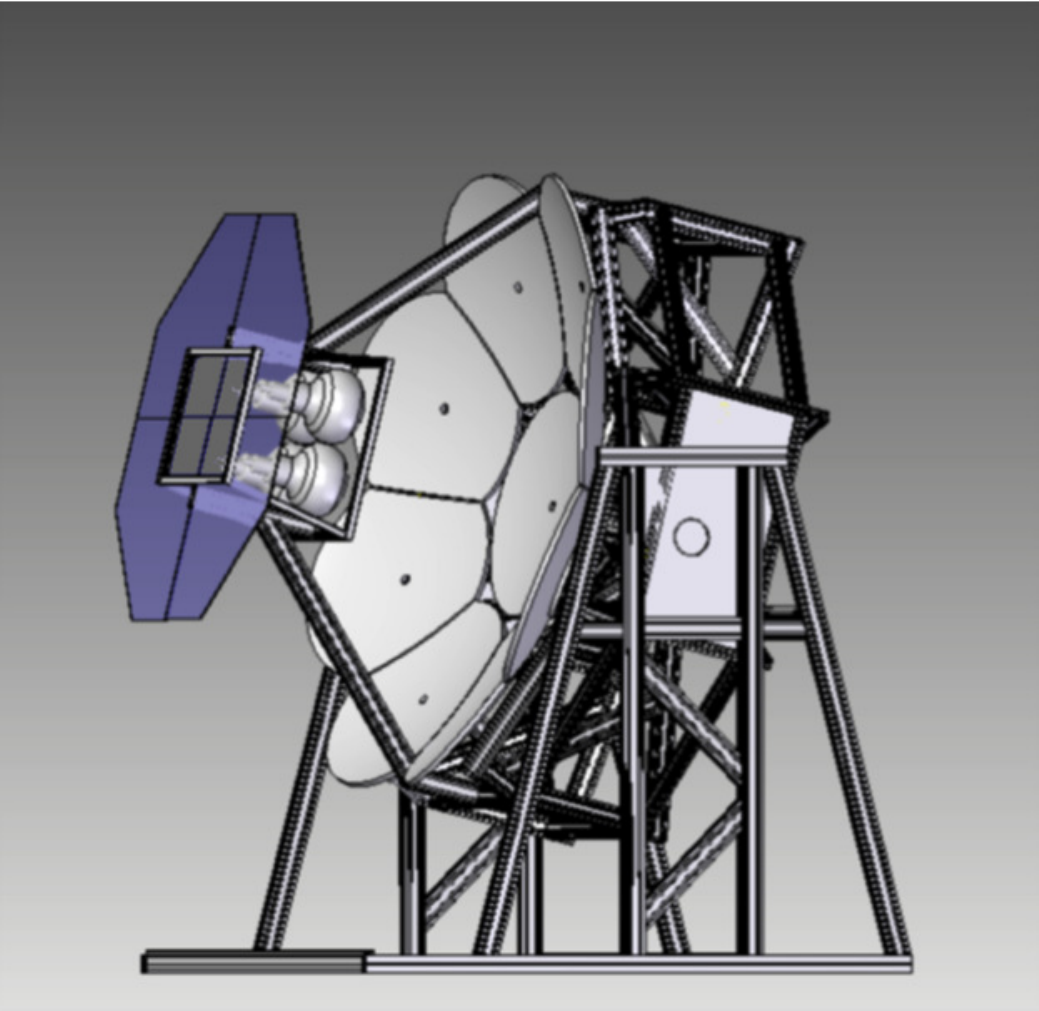}
        \caption{The telescope frame, showing four PMTs at the focus of a 1.6\,m diameter segmented mirror. The support structure is made from aluminium profiles. The UV filter can be seen attached to the periphery of the camera box.}
        \label{fig:fastTel}
    \end{subfigure}%
    ~ 
    \begin{subfigure}[t]{0.5\textwidth}
        \centering
        \includegraphics[width=1.\linewidth]{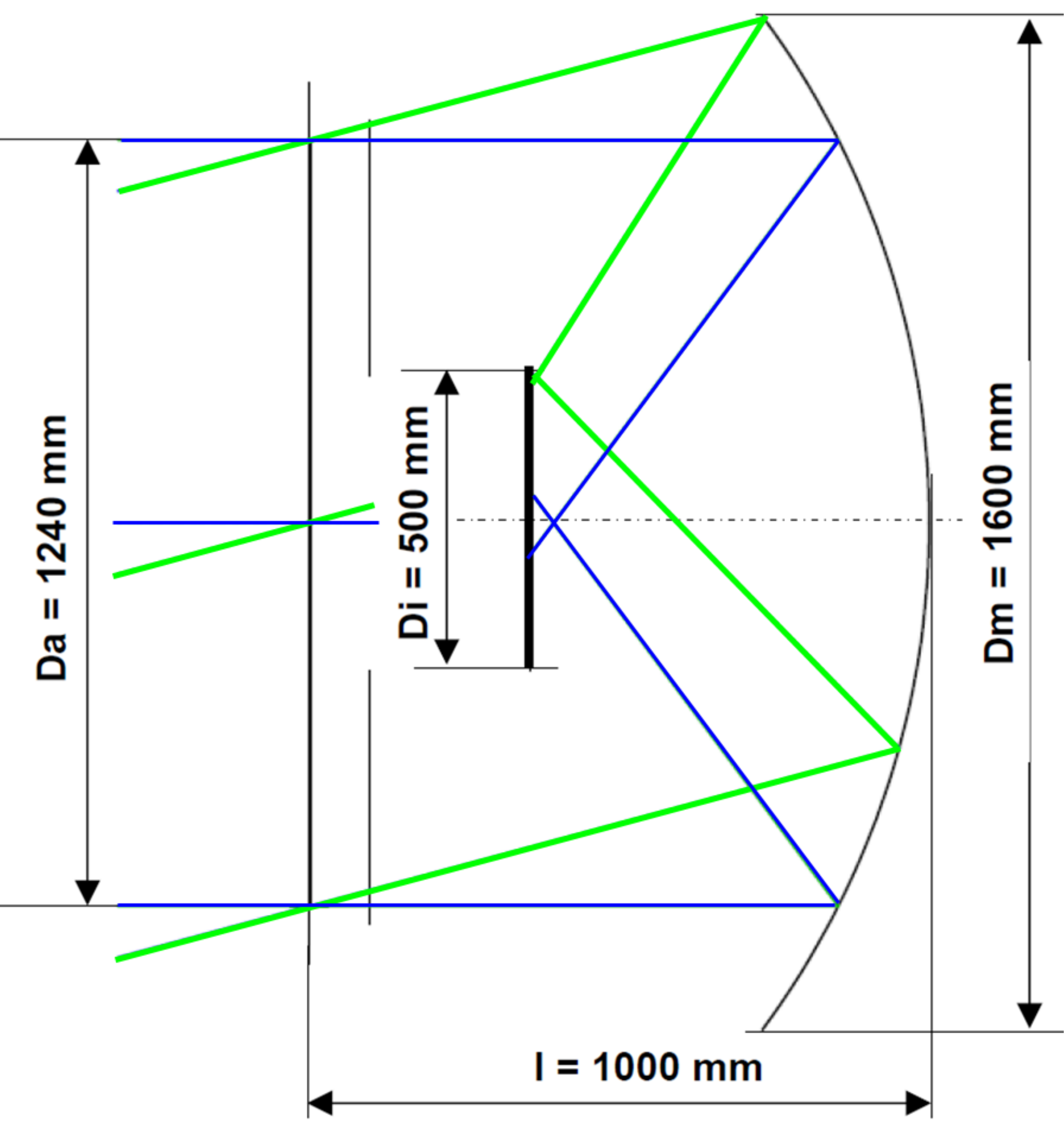}
        \caption{The dimensions of the FAST prototype telescope's optical system. $D_{\textrm{a}}$ is the diameter of the telescope aperture, $D_{\textrm{i}}$ is the side length of the square camera box, $D_{\textrm{m}}$ is the diameter of the primary mirror, and $l$ is the mirror-aperture distance.}
        \label{fig:opt_des}
    \end{subfigure}
    \caption{The mechanical and optical design of the full-scale FAST prototype telescopes.}
\end{figure}

The dimensions of the FAST prototype telescope are shown in Fig.~\ref{fig:opt_des}. An octagonal aperture of height $1.24$\,m is located at a distance of 1\,m from a 1.6\,m diameter segmented spherical mirror (radius of curvature $\sim1.38$\,m). The design fulfils the basic FAST prototype requirements, with an effective collecting area of 1\,m$^2$ after accounting for the camera shadow, and a field-of-view of $30^{\circ} \times 30^{\circ}$.

FAST's central circular mirror and $8$ side mirrors, or ``petals'', are produced by the Joint Laboratory of Optics of the Palacky University and the Institute of Physics of the Academy of Sciences of the Czech Republic from a borosilicate glass substrate. The reflective surface consists of several vacuum coated Al and SiO$_{2}$ layers, offering a relatively constant reflectivity over the fluorescence wavelength band between 300\,nm and 420\,nm, as shown in Fig.~\ref{fig:telEff}.

A UV filter is installed at the aperture of the telescope to reduce the exposure to night-sky background light by blocking photons of wavelength $>400$\,nm. The maximum angle of incidence of the light passing through the filter is $\sim15^{\circ}$, with Fresnel losses being negligible compared to the losses which would be present if the filter were installed on the telescope camera (the maximum incidence angle of light on the filter is $\sim60^{\circ}$ in this configuration). In the latter case, an additional glass window would need to be installed at the aperture to protect the telescope against the environment, resulting in further transmission losses. In addition to reducing the camera's exposure to night-sky background light, the UV filter serves as a protective window against dust and aerosols. We use a ZWB3 filter manufactured by Shijiazhuang Zeyuan Optics. Its spectral transmission is shown in Fig.~\ref{fig:telEff}. The filter is constructed from a number of small segments in order to fit the FAST prototype's octagonal aperture. The individual segments are tessellated using brass ``U'' and ``H'' profiles, resulting in an aperture of area 1\,m$^{2}$.

The telescope's mechanical support structure, shown in Fig.~\ref{fig:fastTel}, is based on commercially available modular aluminium profiles, providing an extremely stable and rigid platform on which the FAST optical system can be mounted. Their light weight allows for easy and inexpensive packaging and transportation, while their modular design makes assembly straightforward. The mechanics consists of a primary mirror stand mounted with a single degree of freedom to facilitate adjustment of the telescope's elevation (the elevation can be set to discrete values of 0$^{\circ}$, 15$^{\circ}$, 30$^{\circ}$ and 45$^{\circ}$ above the horizon). The square camera box (side length 500\,mm), which holds four 200\,mm PMTs, is mounted on a support structure connected to the perimeter of the mirror dish which also holds the octagonal filter aperture. Surrounding the camera box are four flat side mirrors (area $\sim66$\,cm$^2$ mounted at $\sim80^{\circ}$ to the camera surface) designed to reflect light lost due to the enlarged spot size at the camera edges back into the PMTs (see Section~\ref{sec:perf}). The mirror stand contains 9 mirror mounts, each with 2 degrees of freedom to allow for mirror segment alignment. The whole mechanical construction is covered with a shroud to protect the optical system from the surrounding environment.

\subsection{Optical performance}
\label{sec:perf}

\begin{figure}[htb]
	\centering
	\includegraphics[width=1.0\linewidth]{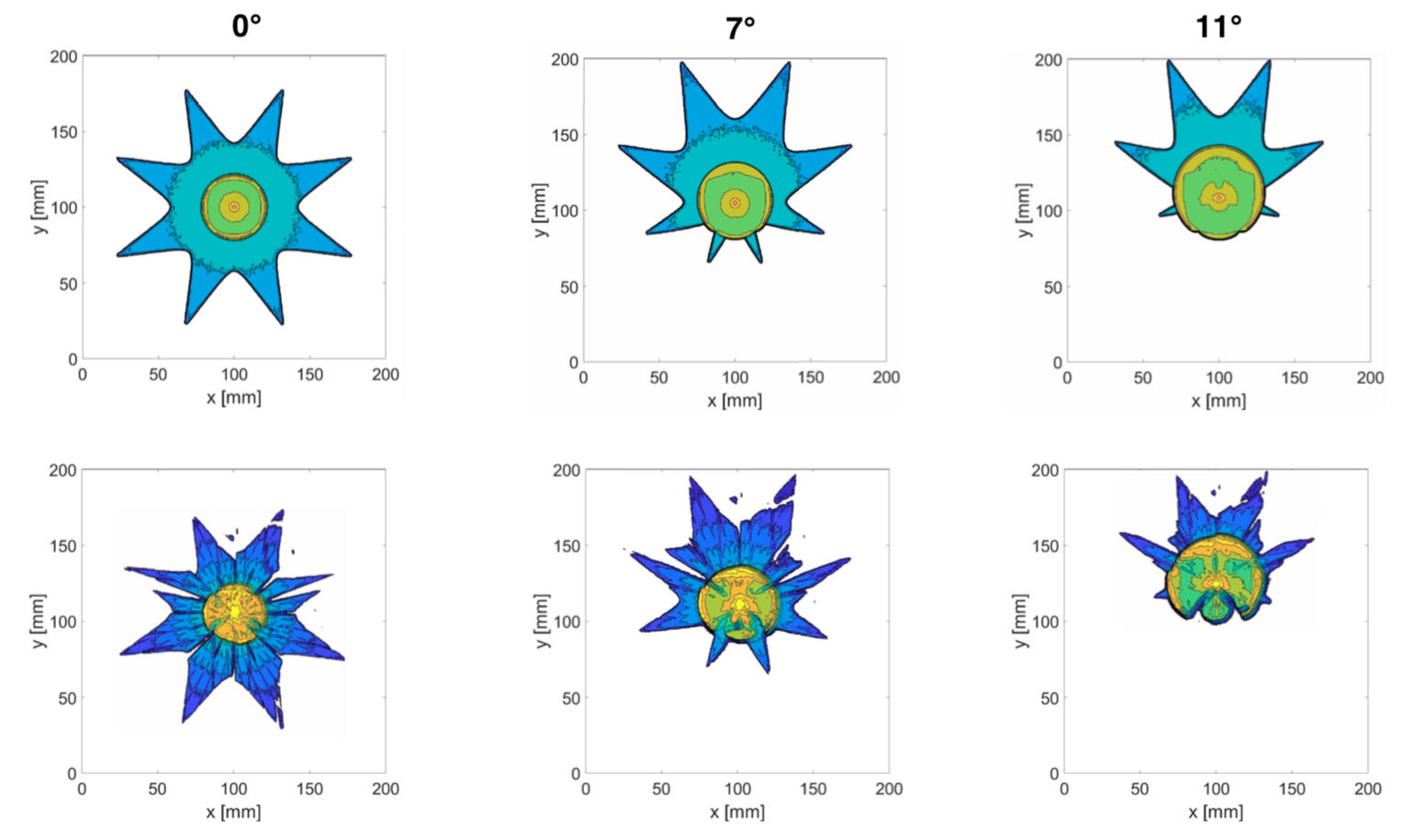}
	\caption{The simulated point spread function (top) for on-axis, 7$^{\circ}$, and 11$^{\circ}$ incidence angles and the true point spread function of a FAST prototype telescope (bottom) measured \textit{in-situ} at the installation site. 
	}
	\label{fig:PSF}
\end{figure}

The top row of Fig.~\ref{fig:PSF} shows the results of a ray-tracing simulation of collimated optical beams at various angles of incidence (on-axis, $7^\circ$, and $11^\circ$) to the telescope aperture performed with the Zemax software package~\footnote{https://www.zemax.com}, with the 200\,mm scale representing the diameter of the PMTs installed in a custom-built box close to the telescope's focal surface. 
The ``star" shape of the optical spot is a result of the octagonal aperture dimensions. In order to minimise the dead space between PMTs, the image plane is moved 25\,mm closer to the mirror (relative to the focal surface) in the prototype design. This serves to eliminate a complete loss of signal for on-axis optical beams where light is focused towards the central point between all four PMTs, by enlarging the optical spot. Some of this signal loss is also mitigated in the prototype design by applying a Tyvec diffusing material to the surface of the camera box between the PMTs.

The lower row of Fig.~\ref{fig:PSF} shows the results of an \textit{in-situ} measurement of the optical point spread function of one of the three prototype telescopes installed at the Telescope Array experiment (see Section~\ref{sec:TA_FAST}). These measurements were made using a point-like light source located at a distance of $\sim150$\,m from the telescope, imaged on a flat screen mounted to the front of the camera box. These measurements show excellent agreement with simulations, verifying not only the performance of the optical system (and the applicability of simulations in assessing its performance), but also the directional alignment of the telescope. The finer structure present in the measured point spread functions is due to the presence of a low chain-link fence between the light source and the telescope.

\begin{figure}
    \centering
    \begin{subfigure}[t]{0.5\textwidth}
        \centering
        \includegraphics[width=1.\linewidth]{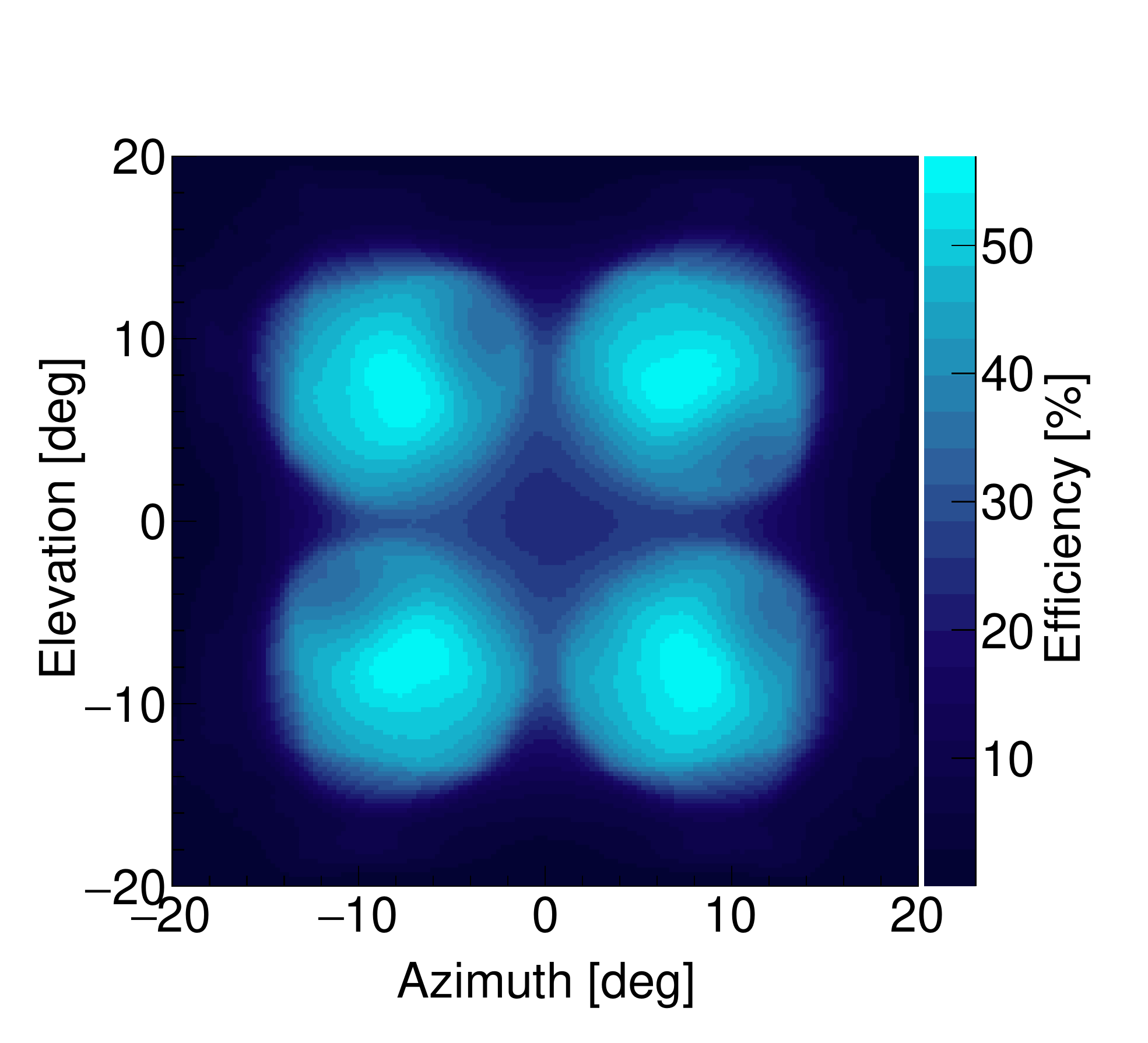}
        \caption{Spectrally-independent directional efficiency of a FAST telescope. See text for details.}
        \label{fig:telEffSE}
    \end{subfigure}%
	~
	\begin{subfigure}[t]{0.5\textwidth}
        \centering
        
        \includegraphics[width=1.\linewidth]{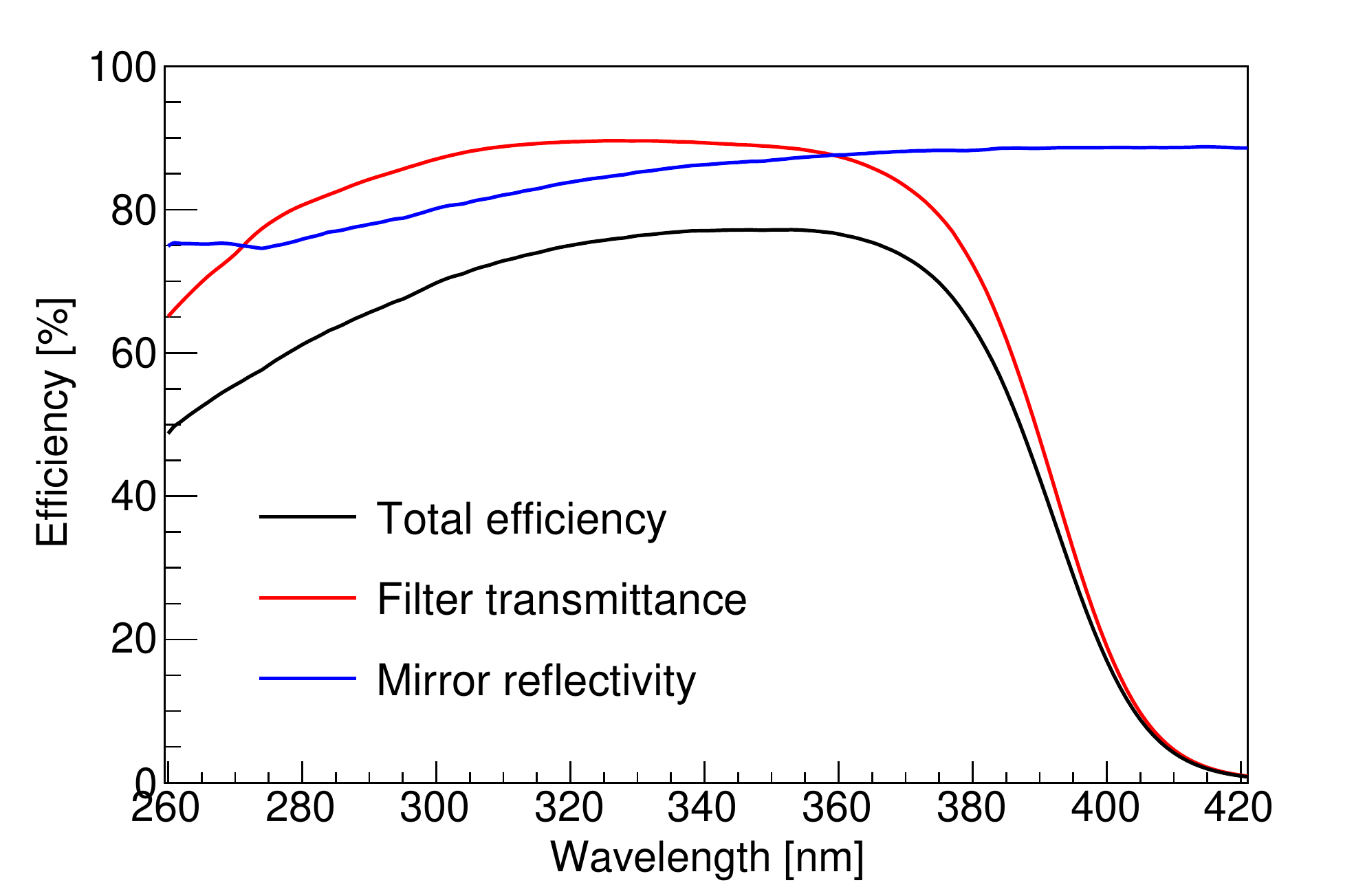}
        \caption{Spectral efficiency of the FAST prototype mirror and UV filter over the fluorescence wavelength range.}
        
        \label{fig:telEff}
    \end{subfigure}
    \caption{Directional and spectral efficiency of the FAST prototype design.}
\end{figure}

Fig.~\ref{fig:telEffSE} shows the results of a full wavelength-independent ray-tracing simulation of the FAST prototype optical system produced using the Zemax software, where the axes represent the angular distance of a collimated beam to the optical axis of the telescope. The simulation model includes a mirror with a realistic surface shape and spectral reflectance (taken from measurements), a complete description of the telescope structure, including the aperture with the filter support structure, the camera box containing the four 200\,mm PMTs, the diffuser attached to the dead space between PMTs, and the four small side mirrors attached to the periphery of the camera. The analysis includes the Fresnel losses on the glass surface of the PMTs. These losses significantly influence the simulation results due to the high incidence angles of light on the hemispherical photocathode surfaces. In addition, the spatially-dependent collection efficiency of the PMTs is taken into account using measurements made using a dedicated set-up at Chiba University~\cite{bib:chibacalib}. This non-uniform collection efficiency across the PMT photocathode arises due to the distance between the photocathode and the first dynode within large-format PMTs, and manifests itself primarily as a ``cold spot" of $\sim25$\% lower efficiency diametrically opposite the first dynode. Note that this non-uniformity factor is included in the simulation.

\subsection{Data acquisition system}
\label{sec:daq}

The FAST camera and electronics chain is comprised entirely of commercially-available components. The camera comprises four AC-coupled 200\,mm PMTs (mod. R5912-03, Hamamatsu), having 8 dynodes each and a maximum operating voltage of $\sim2600$\,V, with active bases (mod. E7694-01, Hamamatsu) arranged in a $2 \times 2$ matrix, and covering a $\sim 30^\circ \times 30^\circ$ field of view (see Fig.~\ref{fig:telEffSE}).


All PMTs were tested in the laboratory at the University of Chicago, where their detection efficiency and differential linearity were measured, and their nominal operating voltages were determined (typically $\sim900$\,V with a positive polarity for a target gain of 5\,$\times$\,$10^4$). The calibration procedure was almost entirely automated, requiring only that an operator successively install each PMT in the light-tight test box, and ensuring reproducible test conditions for the measurement of each of the PMTs. 

\begin{figure}
  \centering
  \includegraphics[width=0.7\linewidth]{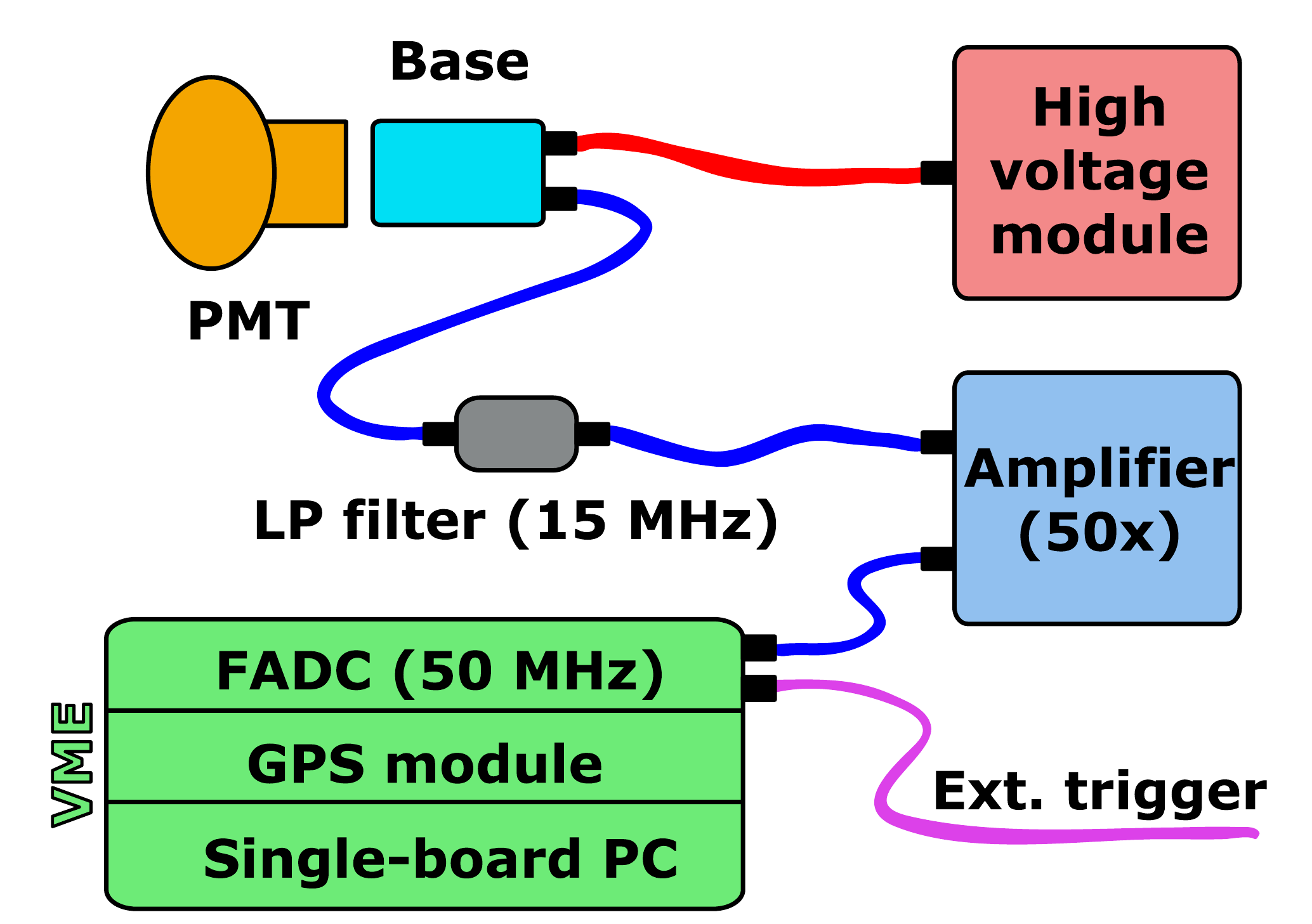}
  \caption{Schematic showing the FAST electronics chain for a single PMT.}
  \label{fig:elec}
\end{figure}

A NIM-mounted module (mod. N1470, CAEN) provides high voltage to the four PMTs. The PMT signals are routed through a 15\,MHz low-pass filter (mod. CLPFL-0015, Crystek) to remove high-frequency noise, before being amplified by a factor of 50 using a fast amplifier (mod. 777, Phillips Scientific). The resultant amplified signal is digitised at a 50\,MHz sampling rate using a 16-channel, 14-bit FADC (mod. SIS3316, Struck Innovative Systeme) hosted in a portable VME crate along with a GPS module (mod. GPS2092, Hytec) providing event time stamps, and a single-board PC (mod. V97865, GE Intelligent platforms) running the DAQ software. Triggers can be provided to the FADC either externally via a NIM pulse input, or internally via a high-threshold internal trigger implemented in the DAQ software. A schematic of the FAST back-end and data acquisition electronics for a single PMT is shown in Figure~\ref{fig:elec}. 

The total cost of a single FAST telescope, including the optical system, mechanical structure, and the electronics and data acquisition system is $\sim$\$25k US. Future FAST prototype iterations will include a custom-designed FPGA-based data acquisition system, including a miniaturised high-voltage supply. The first such prototype is currently under development.

\section{FAST prototypes at the Telescope Array site}
\label{sec:TA_FAST}

\begin{figure}[t]
  \centering
  \includegraphics[width=0.7\linewidth]{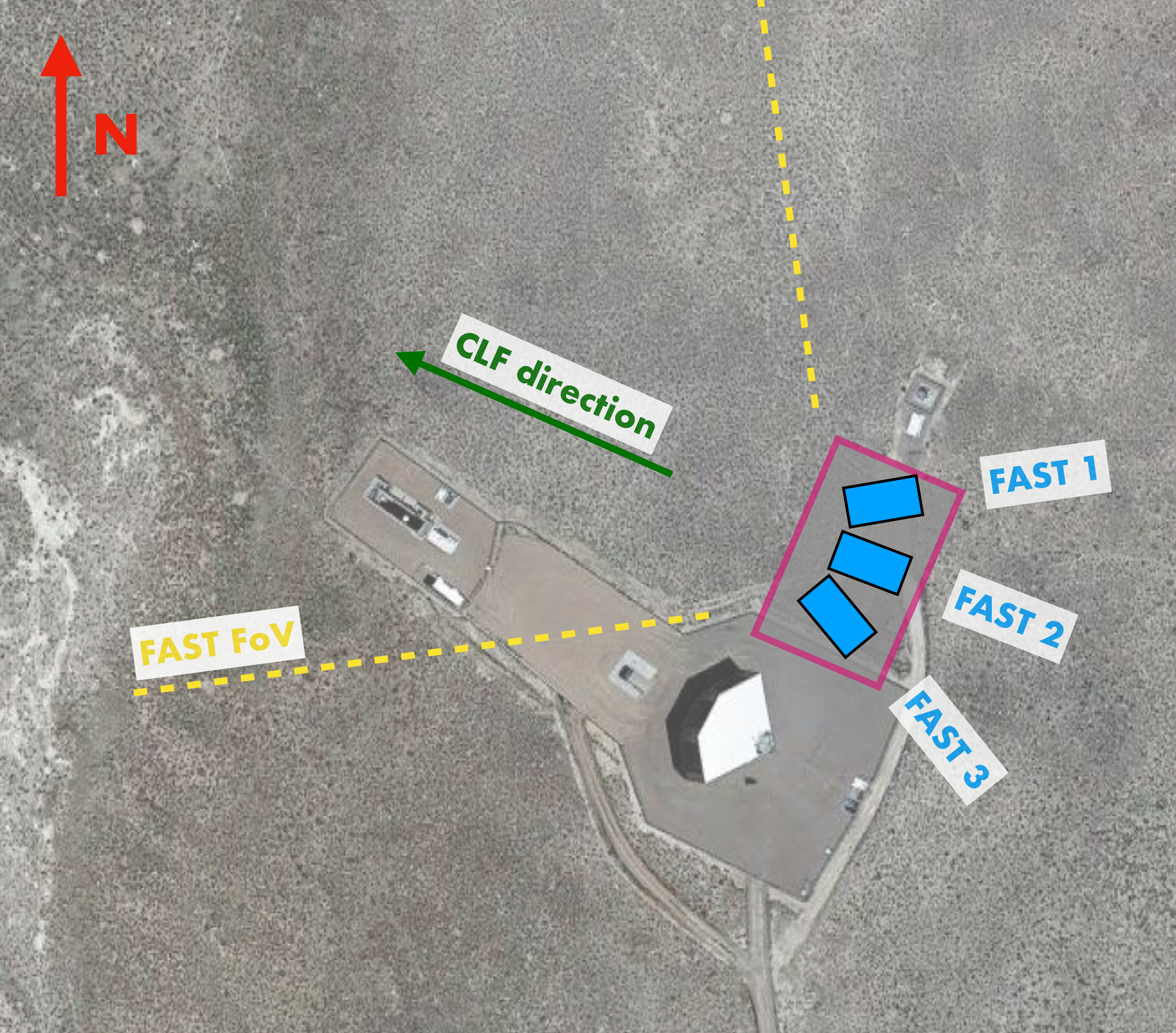}
  \caption{Location and field-of-view (FoV) of the three FAST prototype telescopes installed at the Black Rock Mesa site of the Telescope Array Experiment. The TA fluorescence detector is located south-west of the FAST installation. The central laser facility (CLF) is located $\sim21$\,km away from the BRM site in the indicated direction and is within the FoV of FAST 2.}
  \label{fig:fast_tel}
\end{figure}

\begin{figure}[t]
  \centering
  \includegraphics[width=1\linewidth]{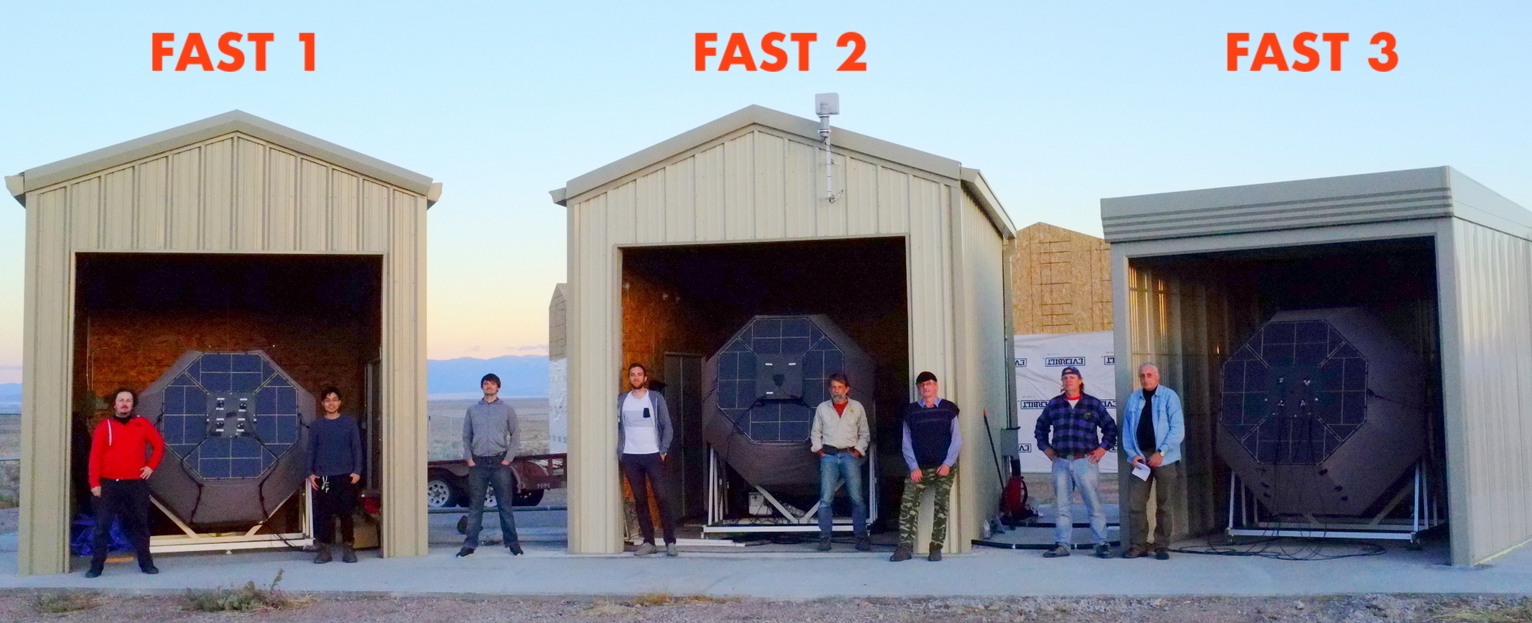}
  \caption{The three FAST prototype telescopes installed at the Black Rock Mesa site of the Telescope Array experiment in central Utah, USA. The FAST All-Sky Camera (FASCam) and Sky Quality Monitor (SQM) can be seen attached to the roof of the central building (see Section~\ref{sec:moni}).}
  \label{fig:fast_photo}
\end{figure}

Following the successful test of the proof-of-concept single pixel FAST telescope at the Telescope Array site in 2015~\cite{bib:firstfast}, a 300\,m$^2$ concrete pad was constructed $\sim50$\,m north of the Black Rock Mesa fluorescence detector to serve as the foundation for three full-size FAST prototype telescopes (see Fig.~\ref{fig:fast_tel} and~\ref{fig:fast_photo}). The site offers access to power and a wireless internet connection, and allows the FAST installation to utilise the external trigger of the adjacent TA fluorescence detector. In addition, the site permits an unobstructed view of TA's vertically-fired 355\,nm ultra-violet laser, the Central Laser Facility (CLF), useful for detector calibration and atmospheric monitoring purposes.

In this section we describe the installation and operation of the first three full-scale prototype FAST telescopes installed at the TA site, along with some preliminary measurements of artificial light sources and cosmic ray air showers made during the first two and a half years of operation.

\subsection{Installation}

In October 2016, September 2017, and October 2018, three full-scale FAST prototypes were installed in dedicated buildings adjacent to the fluorescence detector at the Black Rock Mesa site of the Telescope Array experiment as shown in Fig.~\ref{fig:fast_tel} and~\ref{fig:fast_photo}. 
The combined field of view of the prototypes covers 30$^{\circ}$ in elevation and 90$^{\circ}$ in azimuth, and is fully-contained within the field of view of the TA fluorescence detector.
In each case, the telescope frame was assembled on site, before the PMTs were mounted in the camera box and the UV band-pass filter was installed at the telescope aperture.

The pointing direction of each telescope was calibrated astrometrically using a camera mounted to the exterior of the frame~\cite{bib:fastoptics}. The camera was aligned with the telescope by imaging a distant laser spot directed along the telescope's optical axis. The distance to this imaged laser spot, typically 100 -- 150\,m, defines the parallax in the alignment of the astrometry camera, and provides the dominant uncertainty of $\sim 0.05^{\circ}$ in the telescope alignment. Uncertainties due to the astrometry, which relies on an open-source astrometric calibration service\footnote{http://astrometry.net}, are negligible.

Each of the three buildings is equipped with a remotely-operable shutter to protect the telescope from the environment outside of operating hours. The central building houses the data acquisition electronics, and an area suitable for on-site operation of the telescopes. Two ancillary instruments are installed atop the central hut for monitoring of the night sky: a camera for measurement of the cloud coverage (FASCam) and a sky quality monitor (SQM) for quantifying the night-sky brightness. The operation of these instruments is described in Section~\ref{sec:moni}.



\subsection{Operation and data acquisition}

The FAST prototypes are operated on clear, moonless nights in coincidence with the adjacent Black Rock Mesa fluorescence detector of the Telescope Array experiment. Triggers are received at a typical rate of 3\,Hz, increasing to 10\,Hz every half hour during the firing sequence of the TA CLF, with the external TA trigger being formed when 5 adjacent PMTs measure signals above a pre-determined threshold within a $12.8$\,$\mu$s window~\cite{bib:TAtrigger}.

At the beginning and end of a run, each FAST telescope collects data for 5 seconds with the shutter closed using its high-threshold internal trigger. These data are used to determine the PMT pedestal and to monitor the PMT gain and night-sky background. Additional 3 second measurements of the PMT pedestal are taken periodically throughout the night at 5 minute intervals.

When a trigger is received by the FAST DAQ, a 100\,$\mu$s data-frame sampled at 50\,MHz from each PMT of the three FAST telescopes is recorded. This includes a buffer of $\sim10$\,$\mu$s before the trigger time to allow for an estimation of the pedestal and baseline variance. Data is saved in 5~minute blocks, with a measurement of the pedestal being taken between subsequent DAQ runs using the digitiser's internal trigger.


One of the principal design goals of a FAST telescope is to have it be remotely operable and largely autonomous. The telescopes are operated remotely via SSH connection to a Raspberry Pi single-board computer attached to the DAQ system. The FAST startup and shutdown procedure are fully automated; we use a set of scripts to control and monitor the high voltage, open and close the FAST shutters, perform pre- and post-run pedestal measurements, and collect data throughout the night. An online monitoring page allows remote shifters to monitor the local conditions to determine whether or not it is safe to begin observation, as well as monitor the telescopes throughout the night. Available information includes the instantaneous trigger rate, a measurement of the night-sky brightness, a whole-sky image indicating the presence of clouds, the status of the local storage drives, and live webcam images of the interior of the FAST buildings.

A number of fail-safe systems are in place to protect the FAST telescopes in the event of a power failure, adverse weather conditions, or loss of the remote connection. These include a UPS capable of supplying power while the electronics are shut down and the telescope shutters are closed, and an automatic shutdown routine that commences before sunrise to protect the FAST cameras from high background light.

\subsection{Night-sky background and stability}
\label{sec:nsb}

The average photocathode current $I_{\textrm{p.e.}}$ of the PMTs in a FAST telescope is dominated by the night-sky background (NSB). The NSB decreases after sunset and must be monitored to determine the time at which it is safe to open the telescope shutter and begin data acquisition. The NSB is due primarily to bright stars within the telescope field-of-view and can increase periodically throughout a night of observation as a result of both artificial (e.g. car headlights, aeroplanes, and light pollution) and natural (e.g. lightning) light sources, as well as being affected by cloud coverage and atmospheric transparency (e.g. aerosols). AC coupling of the FAST PMTs does not allow for a direct measurement of the average photocathode current. However, fluctuations in the NSB are recorded as fluctuations in the PMT pedestal, whose variance is linearly related to the average current~\cite{bib:gemmeke2003}.

\begin{figure}
  \centering
  \includegraphics[width=0.7\linewidth]{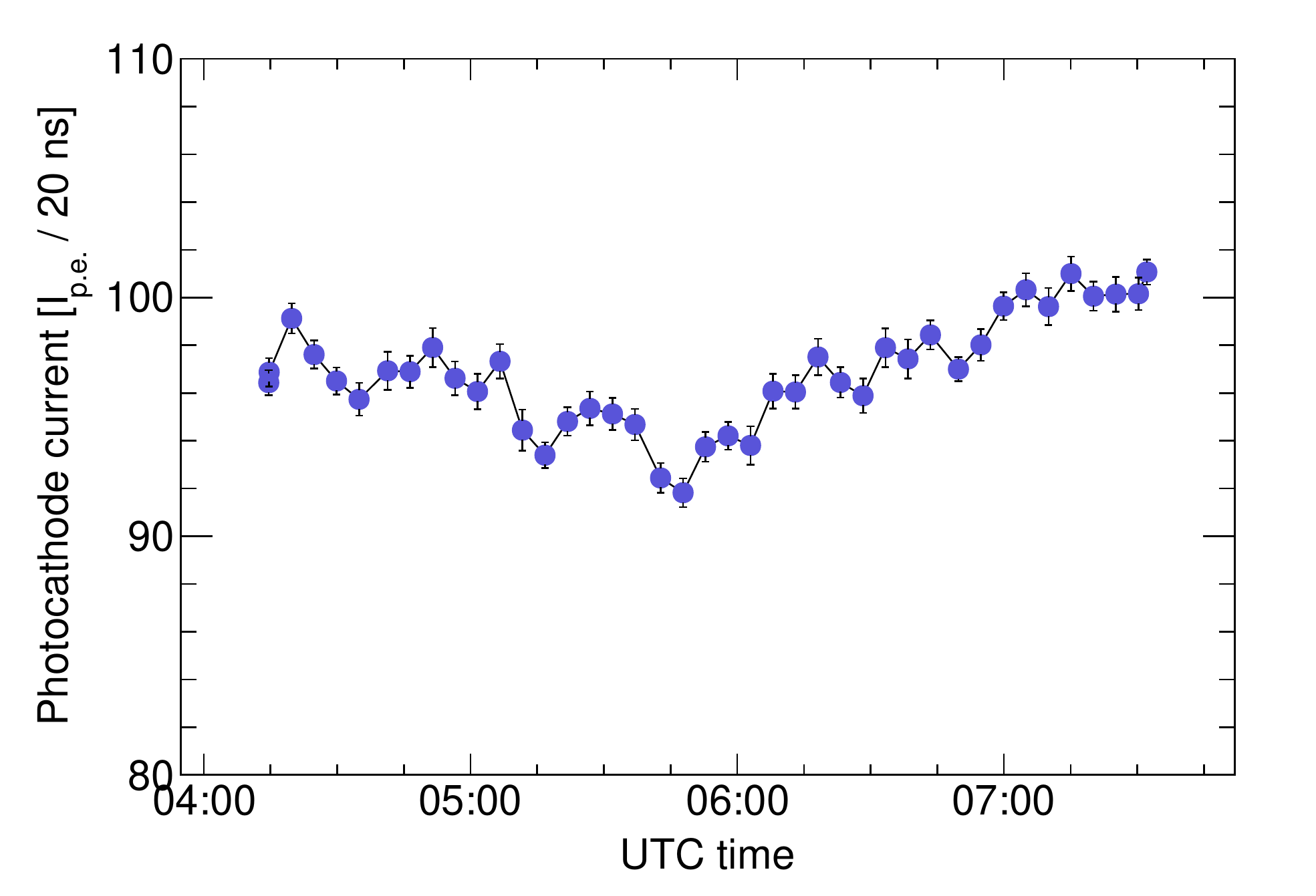}
  \caption{The evolution of the photocathode current during a cloud-free run on Jan. 18$^{\text{th}}$, 2018.}
  \label{fig:nsb_run}
\end{figure}

Two types of pedestal measurements are recorded, first with the shutter closed at the beginning of an observing run, and then at 5 minute intervals during data-taking with the shutter open. With the shutter closed the pedestal fluctuations are dominated by the FAST electronic noise, generating $\sim$13\,p.e.\,/\,20\,ns. With the shutter open the measured photocathode current increases to $\sim$98\,p.e.\,/\,20\,ns, and is dominated by the NSB, indicating that the electronic noise is negligible with respect to the NSB. The evolution of the NSB during a clear-night run of continuous data-taking is shown in Fig.~\ref{fig:nsb_run}.

Two of the 12 FAST PMTs are equipped with a Yttrium-Aluminum-Perovskite (YAP) pulser allowing for a relative calibration of the PMT signal~\cite{Shin:2014uha}. Each YAP pulser is a pulsed UV light source consisting of a Ce-doped YAlO$_{3}$ scintillator crystal paired with a 50\,Bq $^{241}$Am $\alpha$-source. The pulser generates UV photons of peak wavelength 370\,nm with a 20\,ns FWHM pulse width at a rate of $\sim50$\,Hz.

\begin{figure}
    \centering
    \begin{subfigure}[t]{0.5\textwidth}
        \centering
        \includegraphics[width=1.\linewidth]{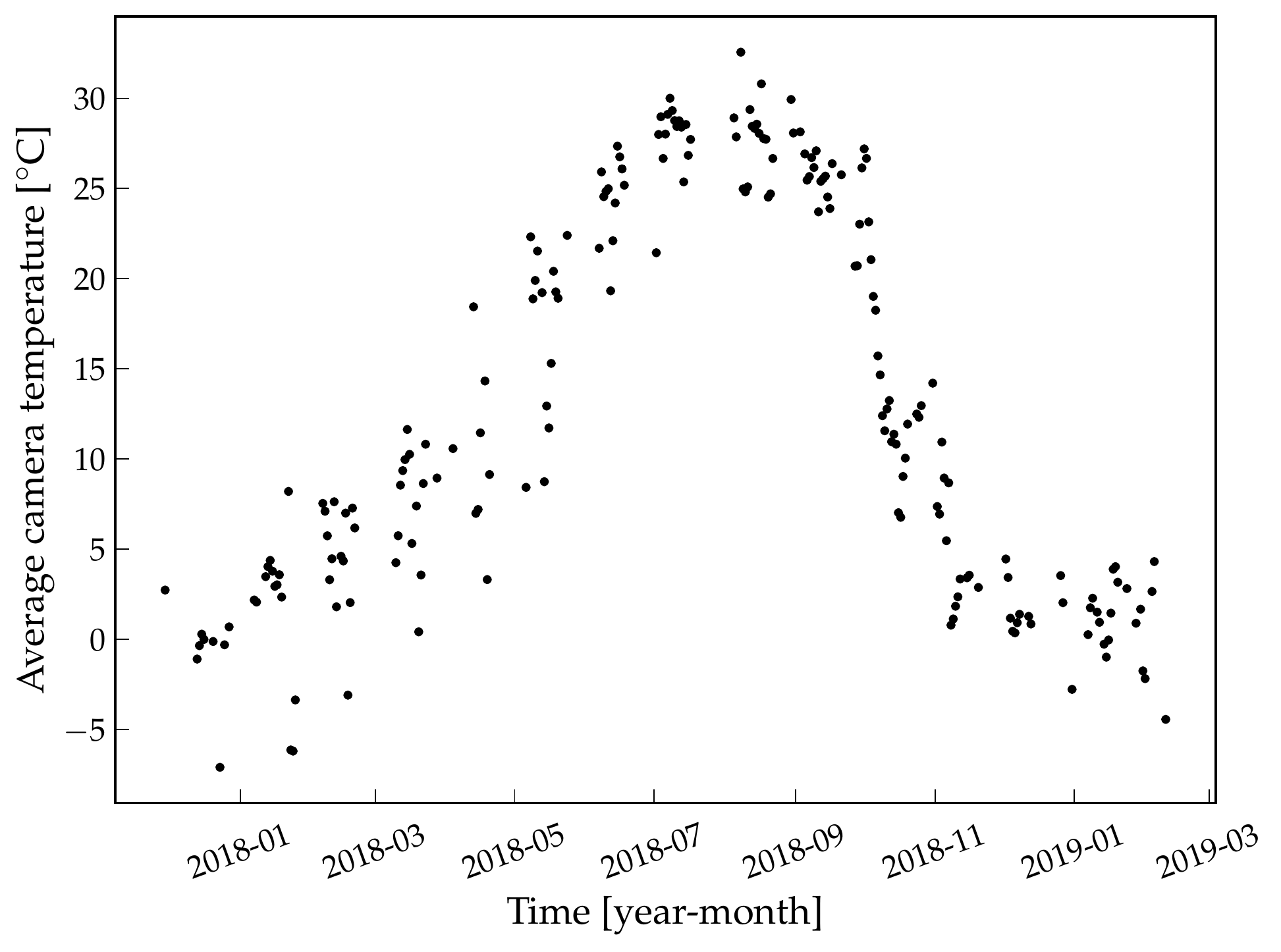}
        \caption{Average temperature inside a FAST camera enclosure at the TA site during each night of observation over a year-long period.}
        \label{fig:avtemp}
    \end{subfigure}%
    ~ 
    \begin{subfigure}[t]{0.5\textwidth}
        \centering
        \includegraphics[width=1.\linewidth]{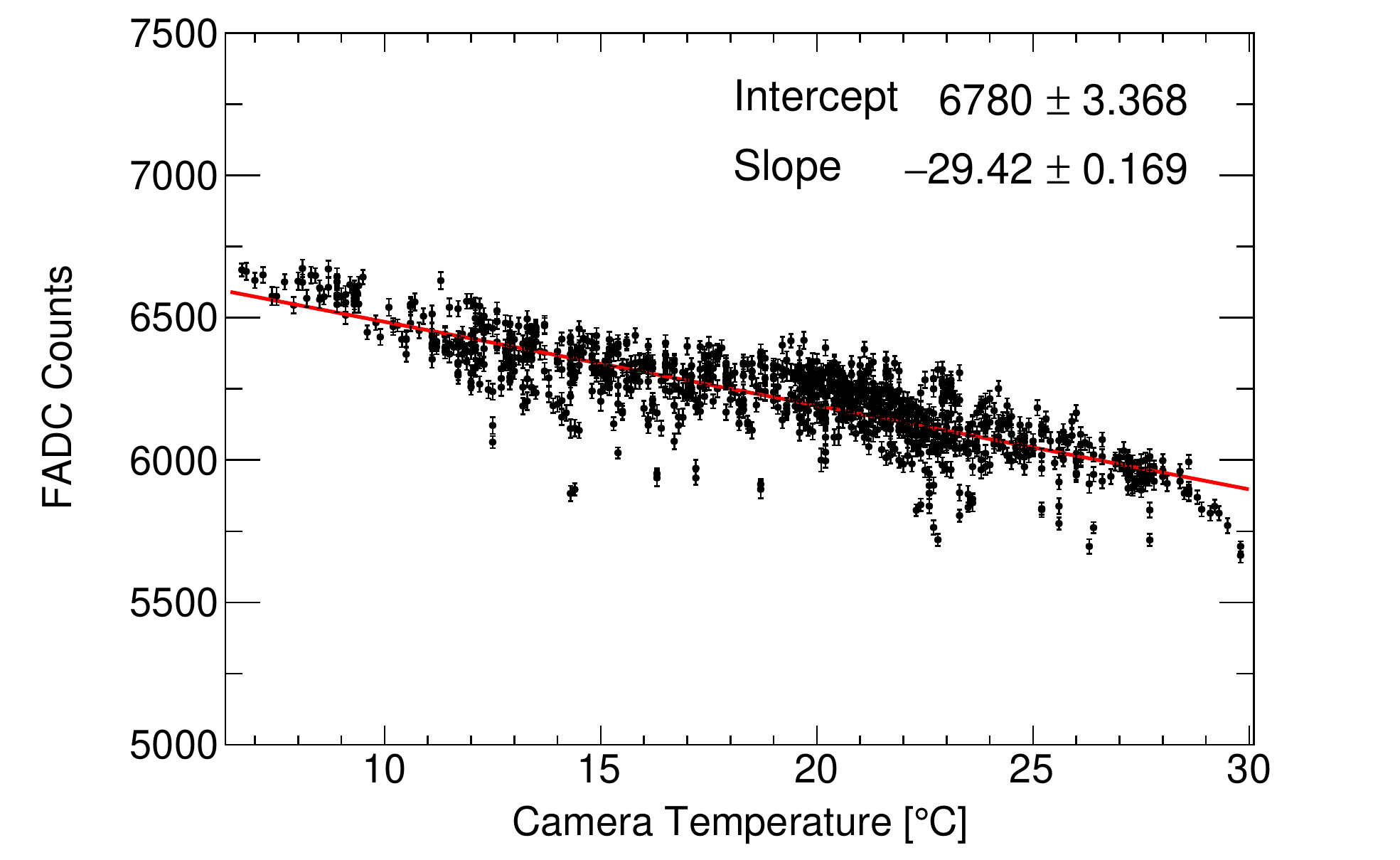}
        \caption{The evolution of the integrated YAP signal as a function of temperature throughout a night of operation.}
        \label{fig:temp_gain}
    \end{subfigure}
    \label{fig:avTemp}
    \caption{}
\end{figure}

Since the FAST telescope buildings are open to the environment during a data-taking run, the operating temperature of the PMTs changes throughout the night, and the average camera temperature changes with the season. As shown in Figure~\ref{fig:avtemp}, there is a 30\,$^{\circ}$C yearly variation in the average temperature inside a FAST camera enclosure at the TA site, and the temperature on a given night can vary by up to 10\,$^{\circ}$C. It is therefore important to track the temperature dependence of the PMT gain for use in later analysis of recorded data.
A YAP pulser measurement is made every 5 minutes during data-taking using a high-threshold internal trigger  for a period of 3\,s, as well as for 5\,s at the beginning and end of every observing night. The integrated YAP signal, in units of FADC counts, is calculated over a 1.1\,$\mu$s region around the signal peak, taking into account the signal pedestal as calculated from the first 500 bins (10\,$\mu$s) of the trace. The evolution of the average integrated YAP signal can be used, in conjunction with temperature measurements made periodically by a sensor attached to the inside of the camera housing in the central FAST telescope, to determine the temperature dependence of the PMT gain. The temperature coefficient was determined to be $-0.434 \pm 0.003$\,\%\,/\,$^{\circ}$C as shown in Figure~\ref{fig:temp_gain}, and is consistent within uncertainties with the manufacturer's quoted specifications.


All YAP pulser and pedestal measurements are stored in an accessible SQL database, along with measurement time stamps and other environmental observables, such as the air and camera housing temperature. This information can later be used for the relative calibration of PMTs during data analysis.


\subsection{In-situ measurement of the filter transmittance and mirror reflectivity}

A possible future array of several hundred fully-autonomous FAST telescopes would cover an enormous ground area, making regular cleaning of each telescope's optical elements unrealistic. It is therefore important to understand and quantify the optical degradation of the individual FAST optical elements following several years of exposure to the Telescope Array's desert environment. A periodic measurement of the optical efficiency of the individual elements can be used as an input parameter for data analyses and is useful in estimating the required cleaning frequency, as well as guiding the design of future prototype iterations.

While each FAST telescope is protected by a shroud, largely shielding the mirror from dust and stray light during hours of operation, the UV filter window is tilted upwards at an angle of $15^{\circ}$ and liable to collect contaminants. These dust particles absorb and scatter light, with a dependence on the particle size and chemical composition. In addition to the build-up of contaminants on the filter, its optical properties can change over time due to degradation of the filter material, such as that caused by exposure to UV light from the sun. The effect of changes in these optical properties is a net loss of optical signal at the telescope camera and a degradation of the optical point spread function.

In October 2018 the filter transmittance and mirror reflectivity of the first two FAST prototypes were measured \textit{in-situ}, using an integrating sphere and a calibrated light source. The relative spectral reflectance of both mirrors was measured using a wide-band fiber-guided deuterium/halogen light source reflected off a small patch of the mirror at an incidence angle of $\sim8^{\circ}$ and into the integrating sphere. The resultant signal was then routed to a spectrophotometer where it was compared to that from a similarly-measured reference surface. The absolute spectral transmittance of the UV filter was measured in a similar way, with the spectral content of the light source being measured using the integrating sphere before placing the UV filter between the integrating sphere and the light source. 

The transmittance of the UV filter was found to have decreased by 5.5\% and 8.5\% over one and two year periods, respectively, due to the build-up of dust and contaminants following remote operation in the field. The decrease in mirror reflectivity over the same time periods was negligible due to the protective shroud shielding the mirrors. 


\subsection{Monitoring of the observing conditions}
\label{sec:moni}

A sound understanding of the atmospheric conditions above the FAST telescopes is essential in interpreting their recorded data. As the atmosphere functions as a calorimeter for FD measurements, its quality is typically one of the largest sources of systematic uncertainty~\cite{bib:EScale}. Atmospheric properties such as the cloud coverage and the presence of micrometer- to millimetre-sized aerosols affect the transmission and scattering of light from developing air showers, and can change significantly over short time-scales. In addition, the height-dependent temperature, pressure and humidity of the atmosphere affect the production of fluorescence light by charged air shower particles, and must be known for an accurate calculation of the fluorescence yield~\cite{bib:airfly2}.

\subsubsection{Aerosols}

The faint fluorescence light produced during the development of an air shower is attenuated on its way to a FAST telescope due to elastic molecular (Rayleigh) and aerosol (Mie) scattering. In addition, strongly forward-beamed Cherenkov light produced by relativistic electrons in the shower can be scattered into the FAST field-of-view. For the highest energy cosmic rays, light from an air shower may have to travel up to 40\,km from its point of emission to a FAST telescope, meaning that the transmission properties of the atmosphere must be well understood. 

Ultra-violet lasers are commonly used for calibration of UHECR fluorescence detectors and for measurement of the transient atmospheric properties within their field-of-view. The Telescope Array experiment features an ultra-violet laser facility, the CLF, operating at $355$\,nm (close to the middle of the atmospheric fluorescence band), which fires 300 vertical laser shots at a rate of 10\,Hz at a nominal energy of $\sim4.4\,\text{mJ}$ (approximately equivalent in intensity to a shower of $10^{19.2}$\,eV) through the field-of-view of the site's fluorescence detectors every half-hour during routine operations~\cite{bib:clf}. 

\begin{figure}[htb]
	\centering
	\includegraphics[width=0.5\linewidth]{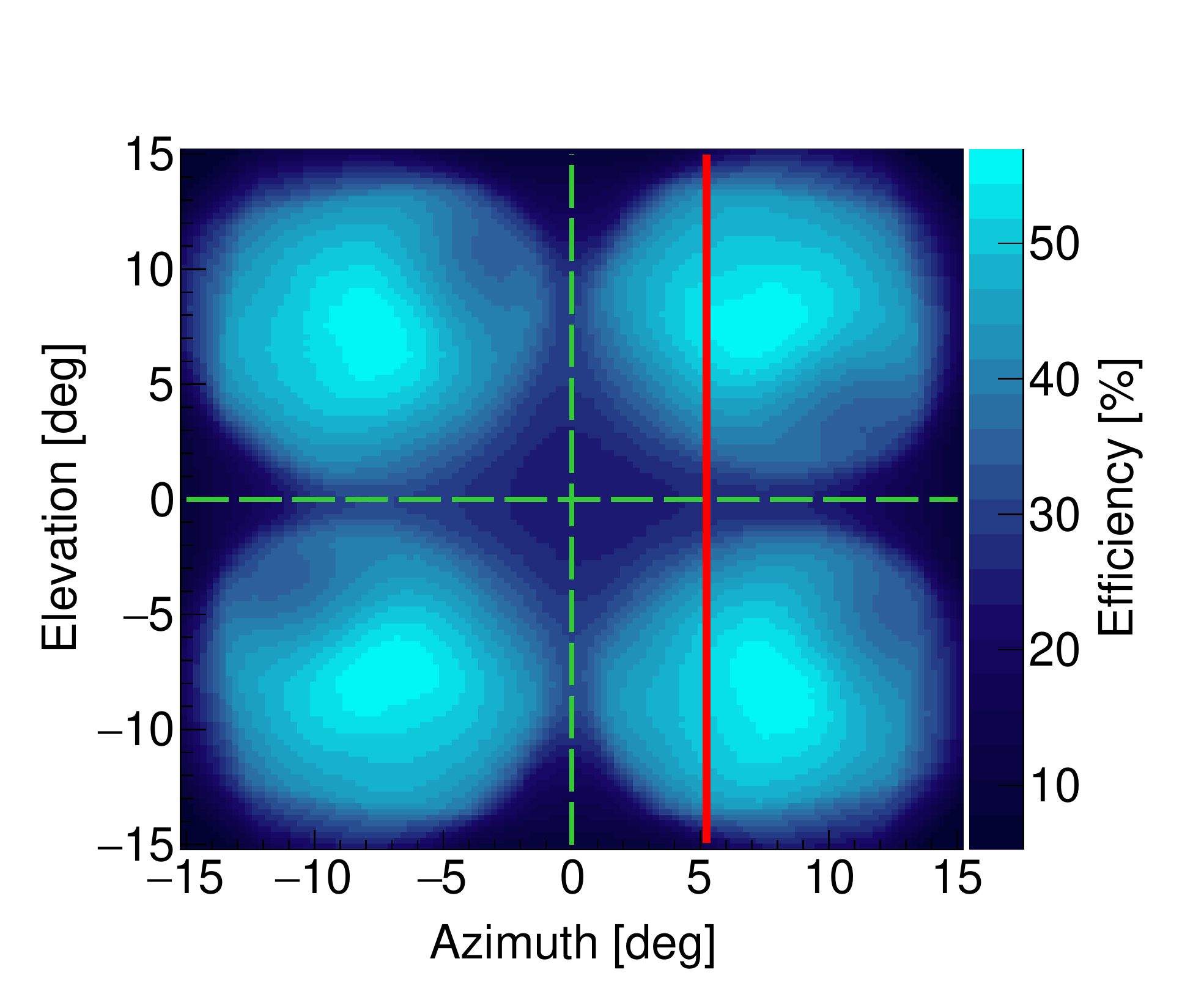}
	\caption{The path of the vertically-fired CLF, shown in red, across the camera of the central FAST telescope. The contours show the directional efficiency of the telescope, including a ray-tracing simulation of the telescope optics, and the azimuthally-dependent response of the PMTs.}
	\label{fig:cameraCLF}
\end{figure}

\begin{figure}[t]
    \centering
    \begin{subfigure}{0.5\textwidth}
        \centering
        \includegraphics[width=1.\linewidth]{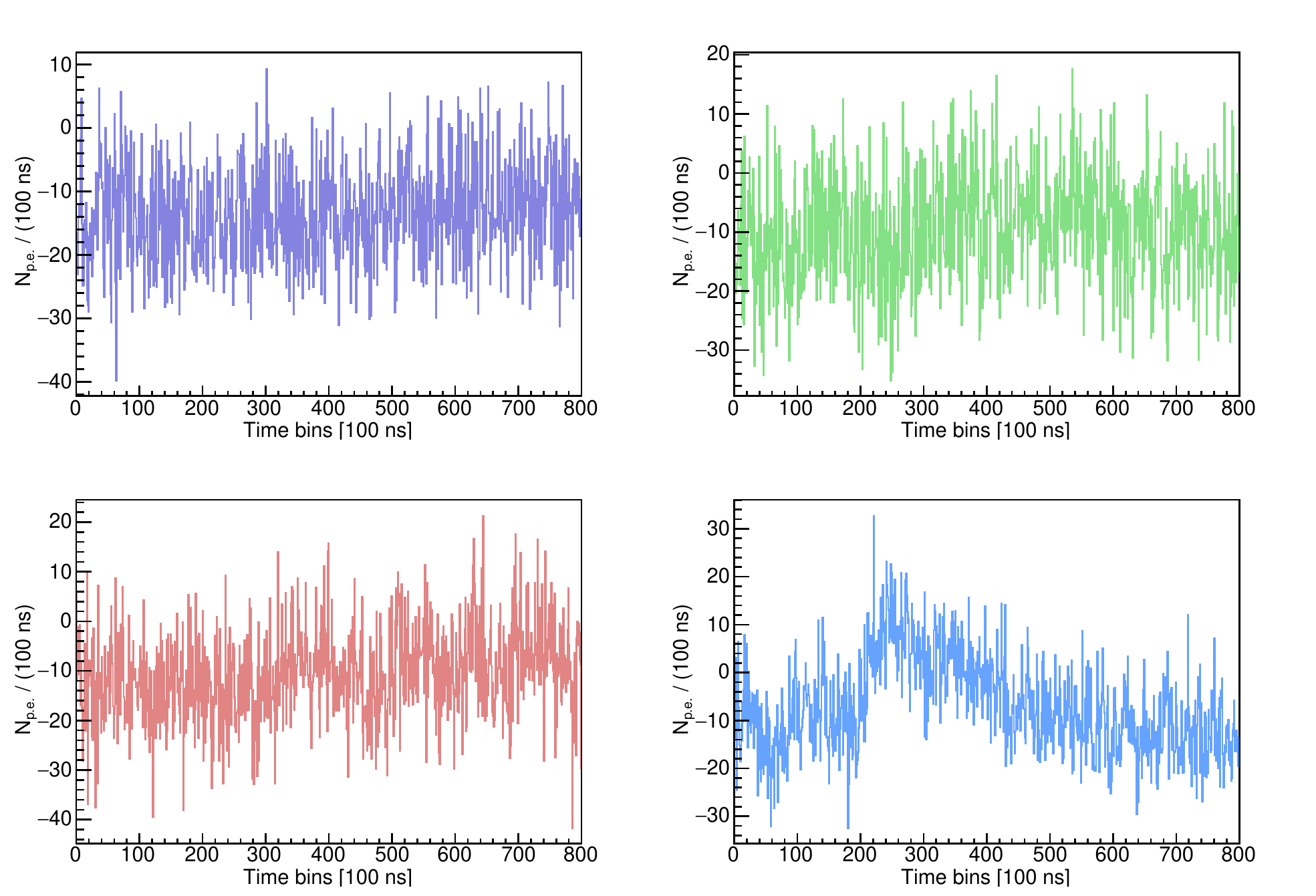}
    \end{subfigure}%
    ~ 
    \begin{subfigure}{0.5\textwidth}
        \centering
        \includegraphics[width=1.\linewidth]{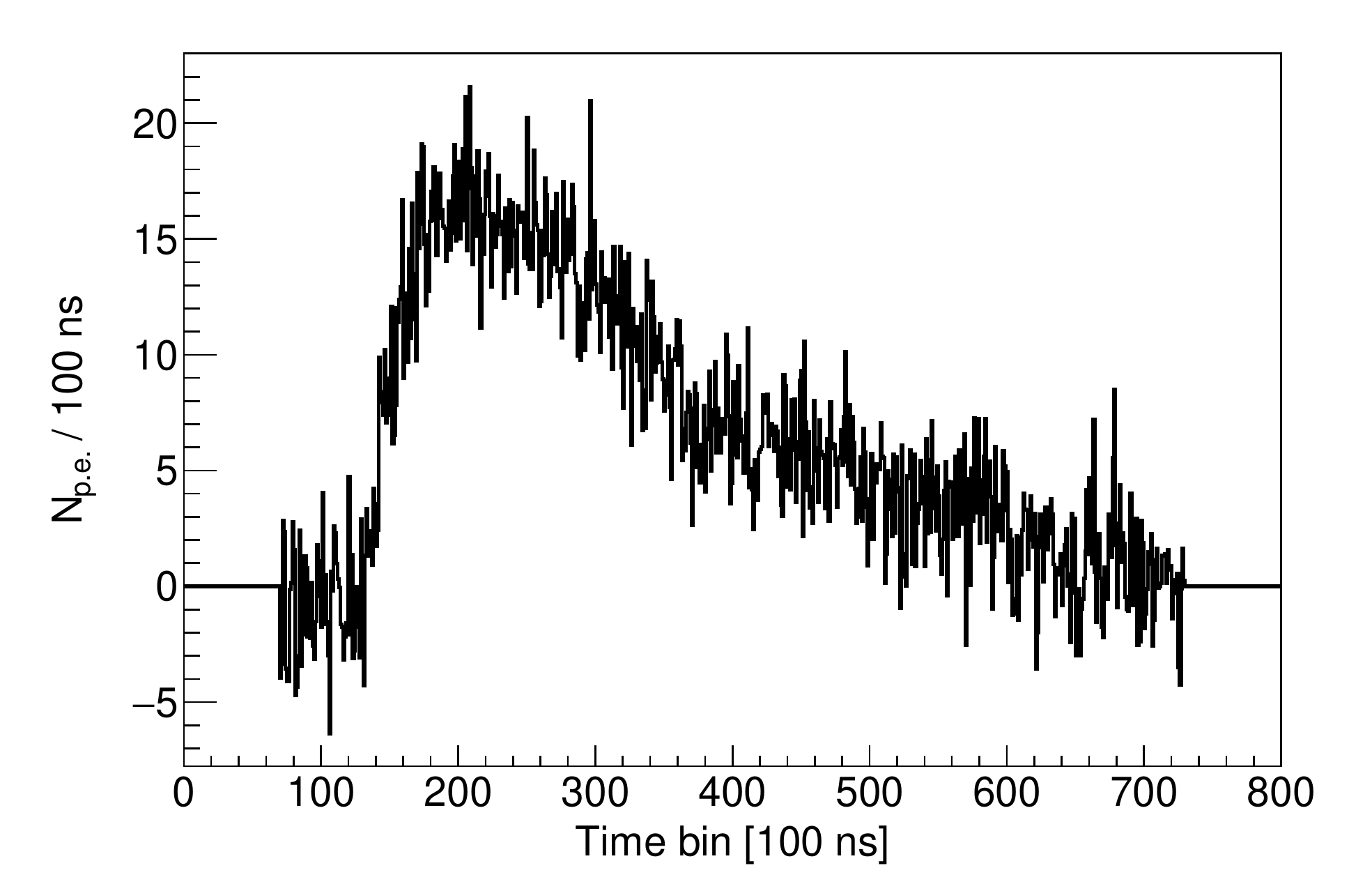}
    \end{subfigure}
    \caption{Left: A single measurement of a vertical CLF laser shot with the central FAST telescope. Right: Average of 200 measurements of the vertically-fired CLF laser, with each individual shot corrected for jitter ($\langle\Delta t\rangle \approx 2.2$\,$\mu$s) in the GPS timing.}
    \label{fig:exCLF}
\end{figure}

The Black Rock Mesa site at which the three FAST telescopes are installed is located approximately 21\,km south-east of the CLF. The alignment of the central FAST telescope was chosen such that the CLF laser track passes directly through two of its pixels (see Fig.~\ref{fig:cameraCLF}), providing a test signal of known source intensity suitable for both calibration and atmospheric monitoring purposes. An example of a TA CLF signal measured by a FAST telescope is shown in Figure~\ref{fig:exCLF}. All vertical laser shots measured by FAST during a firing sequence are corrected for jitter in the GPS timing ($\langle\Delta t\rangle \approx 2.2$\,$\mu$s) and averaged to increase the signal-to-noise ratio. The extraction of the extinction properties of the aerosol atmosphere from these vertical laser traces is described in Section~\ref{sec:atmosanalysis}.

\subsubsection{Cloud coverage and night-sky brightness}

\begin{figure}[t]
    \centering
    \begin{subfigure}{0.5\textwidth}
        \centering
        \includegraphics[width=1.\linewidth]{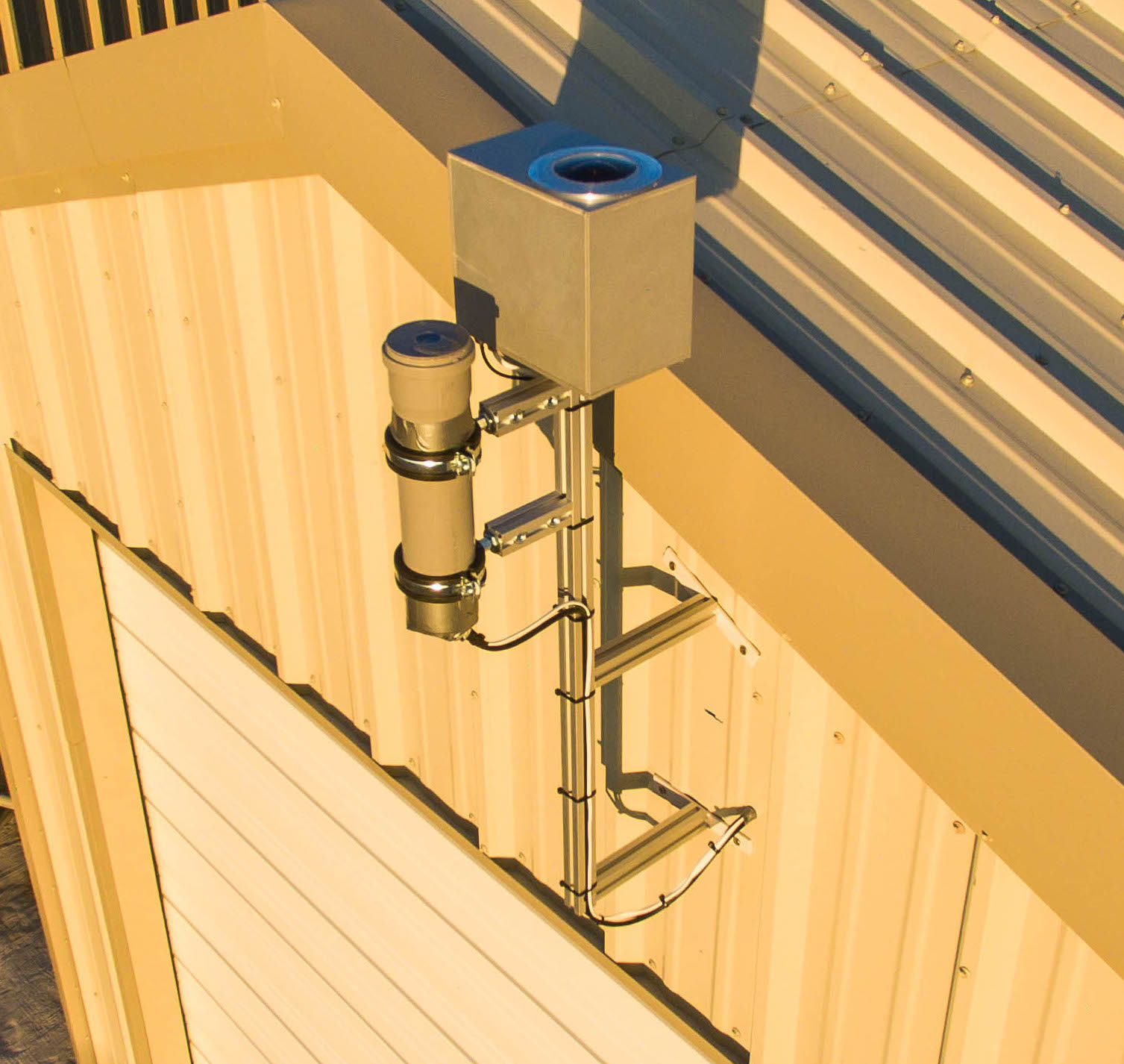}
    \end{subfigure}%
    ~ 
    \begin{subfigure}{0.5\textwidth}
        \centering
        \includegraphics[width=1.\linewidth]{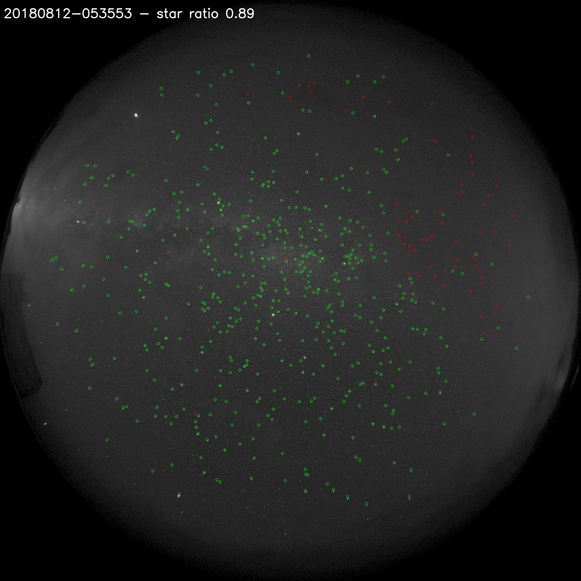}
    \end{subfigure}
    \caption{Left: The FASCam and SQM installed on the central FAST telescope. Right: An example of a FASCam analysis. Identified stars are indicated with green circles, while stars that should be observable but are obscured by cloud are shown as red circles.}
    \label{fig:FASCam}
\end{figure}

Local measurements of the cloud coverage above the FAST telescopes are essential for interpretation of their recorded data. The FASCam, the FAST All-Sky Camera, is a CCD camera with a $180^{\circ}$ field-of-view, equipped with a Moravian Instruments G2-4000 Peltier-cooled KAI-4022 CCD chip and mounted vertically on the exterior roof of the central FAST enclosure (see left of Fig.~\ref{fig:FASCam}). It uses a 5 position adjustable filter wheel with Johnson BVR filters, and a Sigma $4.5\,$mm $f/2.8$ fish-eye lens. The FASCam is controlled via a Raspberry Pi single-board PC mounted to the wall inside the telescope hut. FASCam provides 30\,s exposures of the night sky using the Johnson filters, as well as a 180\,s exposure through the UV filter every 10 minutes during data-taking. An astrometry-based analysis compares images of star positions with known coordinates in each wavelength band from the Tycho-2 catalog and calculates the ratio of the number of visible to observable stars in order to estimate the could coverage. An example of a single FASCam analysis is shown in the right of Fig.~\ref{fig:FASCam}.

Measurements taken over 265 nights of observation since the FASCam was installed in September 2017 suggest a clear sky during 55\% of the measurement time (a clear sky is defined as a visible/observable star fraction $>0.8$). The cloud coverage is continuously monitored during FAST data-taking, and is available to shift operators through a simple web-based monitoring interface. In the case of very cloudy conditions, data-taking can be paused until conditions improve. The cloud fraction is recorded in a database and can be later queried during the reconstruction of air shower data.

Images taken with FASCam's UV filter can be used to characterise the night-sky background within the field of view of individual FAST PMTs (in the case that the sky is free of clouds). As the pointing direction of each pixel of FASCam is calculated using photometry, we are able to map the FASCam pixels onto the FAST PMTs, and hence estimate the NSB signal from the sky exposure in the UV band.

\begin{figure}[htb]
	\centering
	\includegraphics[width=0.7\linewidth]{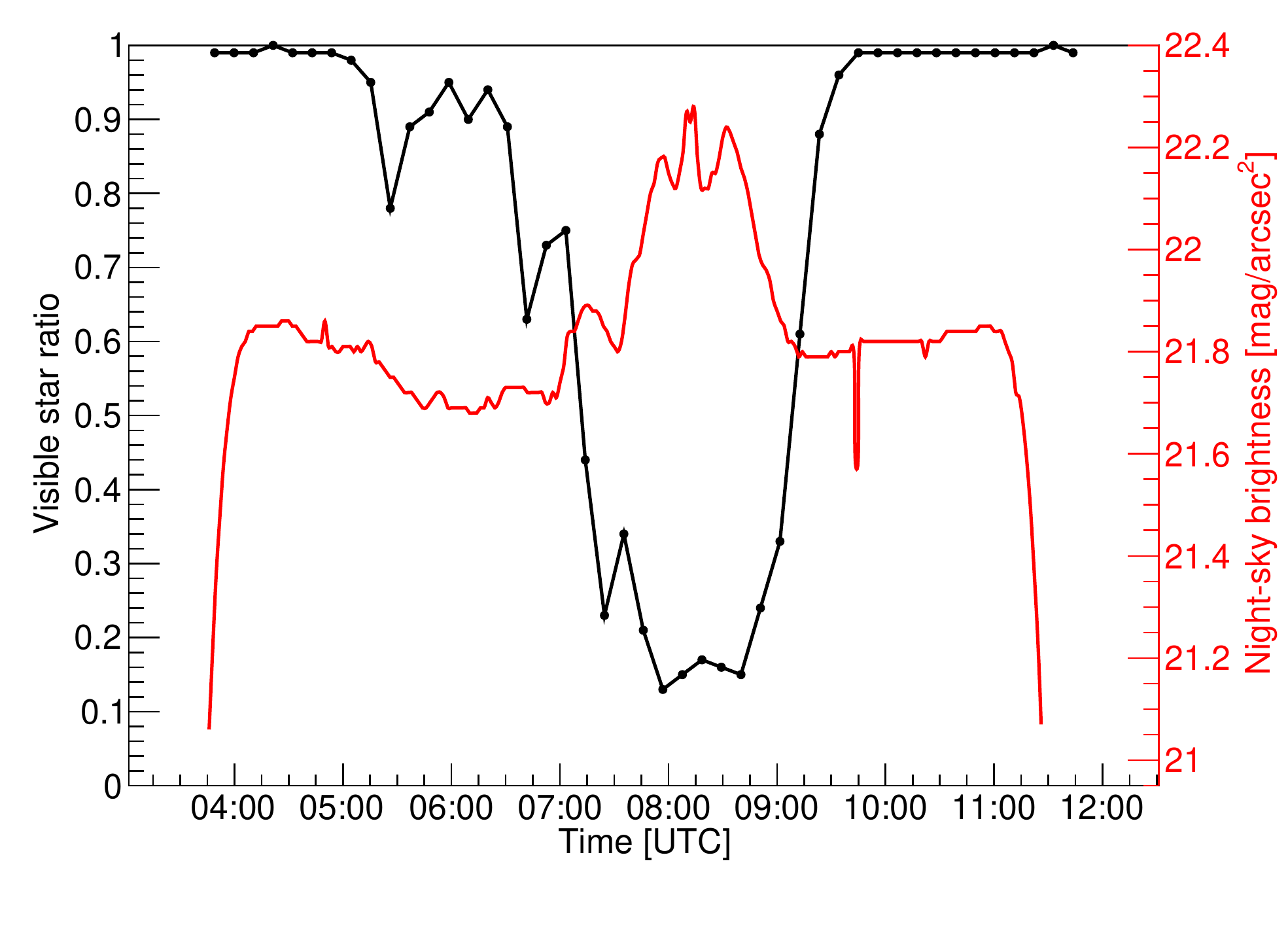}
	\caption{Coincident measurements of the cloud coverage and night sky brightness as recorded by the FASCam and SQM on August 12$^\text{th}$, 2018. The cloud coverage increases to a maximum at around 08:30\,UTC, with a corresponding decrease in the night sky background light measured by the SQM. The anti-correlation between the night-sky brightness and cloud coverage early in the night is likely due to the reflection of the moon on the developing clouds.}
	\label{fig:FASCamSQM}
\end{figure}

The sky-quality monitor (SQM) is a commercial device attached to the roof of the central FAST enclosure, used to measure the night-sky brightness in magnitudes per square arc-second. The device has a precision of $\pm0.1$\,mag\,/\,arcsec$^2$. Measurements taken over the past year suggest a median NSB of 21.6\,mag\,/\,arcsec$^2$, similar to that of the CTA candidate sites~\cite{bib:ctaasc}. Shown in Fig.~\ref{fig:FASCamSQM} are coincident measurements of the cloud coverage and night sky brightness as recorded by the FASCam and SQM on August 12$^\text{th}$, 2018. The cloud coverage increases to a maximum at around 08:30\,UTC, with a corresponding decrease in the night sky background light measured by the SQM.

\subsection{Measurement of air showers}
\label{sec:showers}

The three FAST prototype telescopes have been operating in coincidence with the Black Rock Mesa fluorescence detector of the Telescope Array experiment for a total of 515 hours as of March, 2019. A shower search was performed on $\sim150$ hours of recorded data up to the end of December, 2018, taken from data collection periods free from debugging activities and artificial light source tests, and driven by well reconstructed TA FD events which generated an external trigger for the FAST DAQ.

In addition to shower events and laser pulses from the Telescope Array's CLF, the FAST prototypes are sensitive to various background signals such as airplanes passing through the field of view, lightning, and short time-scale phenomena such as muons coincident with the PMT photocathodes and low energy Cherenkov-dominated showers directed towards the telescopes. To locate shower events in the $\sim 1.37$\,M triggers obtained during this data collection period we applied a finite impulse response (FIR) trapezoidal filter to each recorded trace, looking for excursions above a pre-defined ``event candidate" threshold. Such a filter was chosen for its ability to effectively cancel the high night-sky background level present in FAST data, while also having similarities to the internal triggering algorithm of the DAQ system, allowing for \textit{a-posteriori} tuning of the trigger parameters. For this search, we used two 2.5\,$\mu$s windows with a gap time of 100\,$\mu$s, chosen based on the expected time width of a typical shower signal, and tested on a subsample of observed showers. A number of additional cuts were then applied to remove surviving airplane triggers, CLF shots, and short time-scale muon-like events. The remaining triggers were matched in time with air showers observed in coincidence by the adjacent Telescope Array fluorescence detector.

\begin{figure}
	\centering
	\includegraphics[width=1.\linewidth]{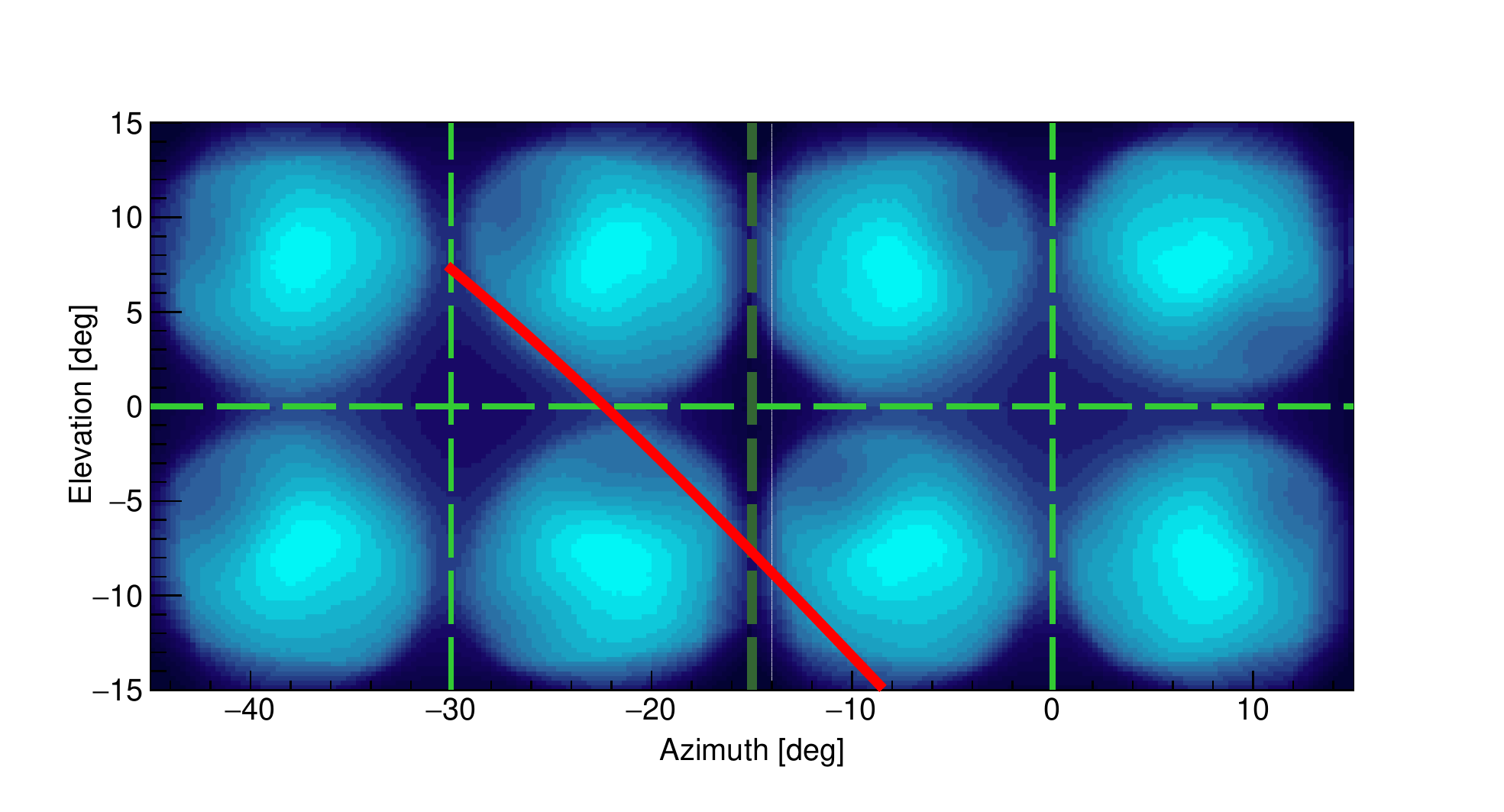}
	\label{fig:shower_track}
		\\
	\includegraphics[width=1.\linewidth]{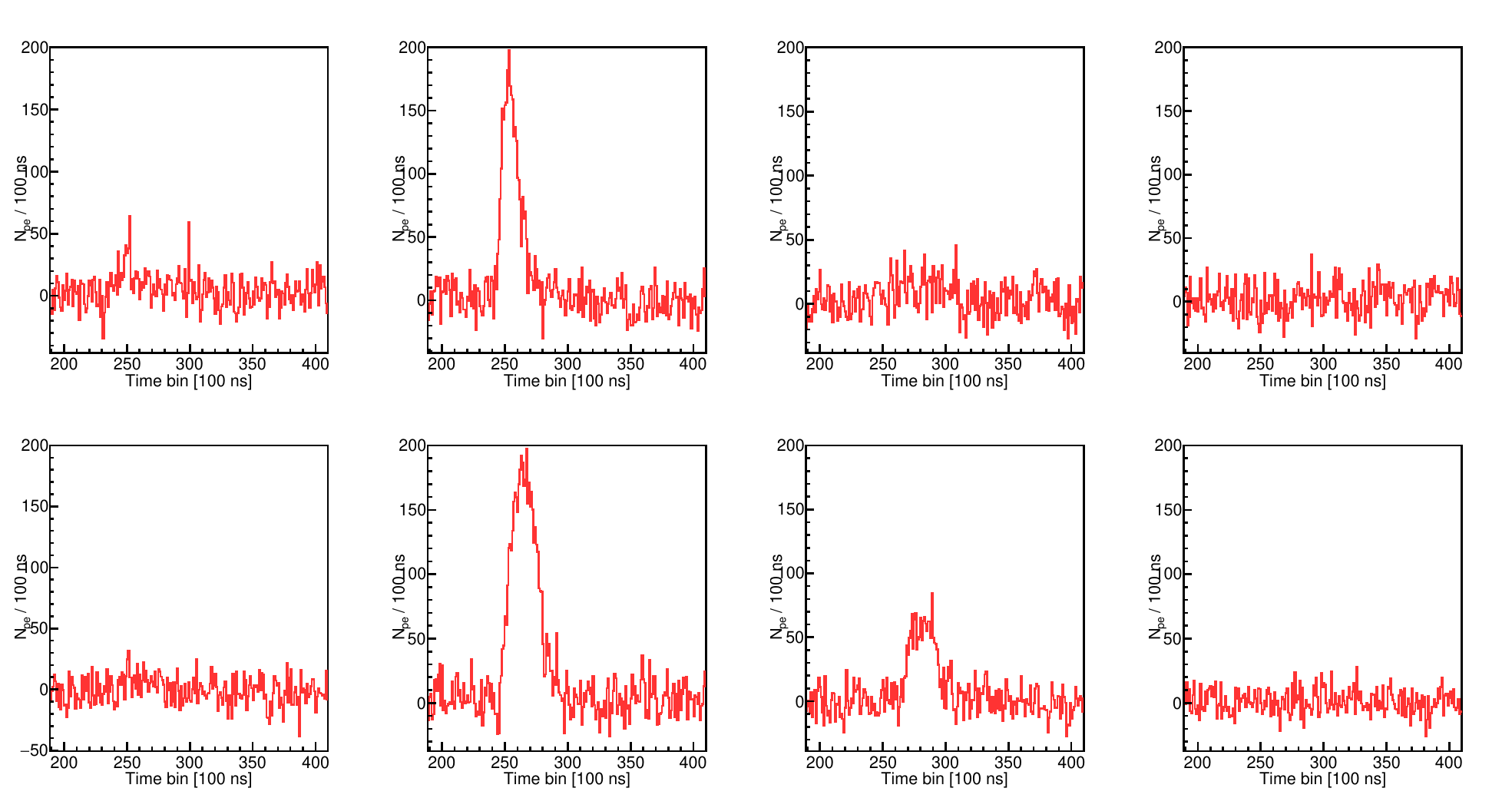}
	\caption{Event measured by FAST 1 and FAST 2 on May 15$^\text{th}$, 2018, with an energy of $\sim19$\,EeV and a zenith angle of $\sim55^{\circ}$. Top: The path of the shower projected onto the FAST focal surface. Bottom: The signal measured in the 8 PMTs of FAST 1 and FAST 2.}
	\label{fig:shower}
\end{figure}

An example of a measured event, the highest energy shower recorded by a FAST prototype thus far, is shown in Fig.~\ref{fig:shower}. The pictured event was observed on May 15$^\text{th}$, 2018, during the operation of the first two prototype telescopes, with an energy of $\sim19$\,EeV and a zenith angle of $\sim55^{\circ}$ (as provided by the Telescope Array monocular reconstruction of the coincident measurement). Shown in the top pane is the geometry of the shower projected onto the FAST focal surface. The bottom pane shows the measured signal in the 8 PMTs of the two telescopes that observed the shower. The time evolution of the event can be seen clearly, and the shape and amplitude of the recorded traces are in good agreement with simulations produced using the best-fit parameters from a top-down reconstruction of the event (see Section~\ref{sec:sim}), further confirming the calibration of the telescope pointing directions, and their spectrally-dependent optical response.

\begin{figure}[t!]
    \centering
    \begin{subfigure}[t]{0.5\textwidth}
        \centering
        \includegraphics[width=1.\linewidth]{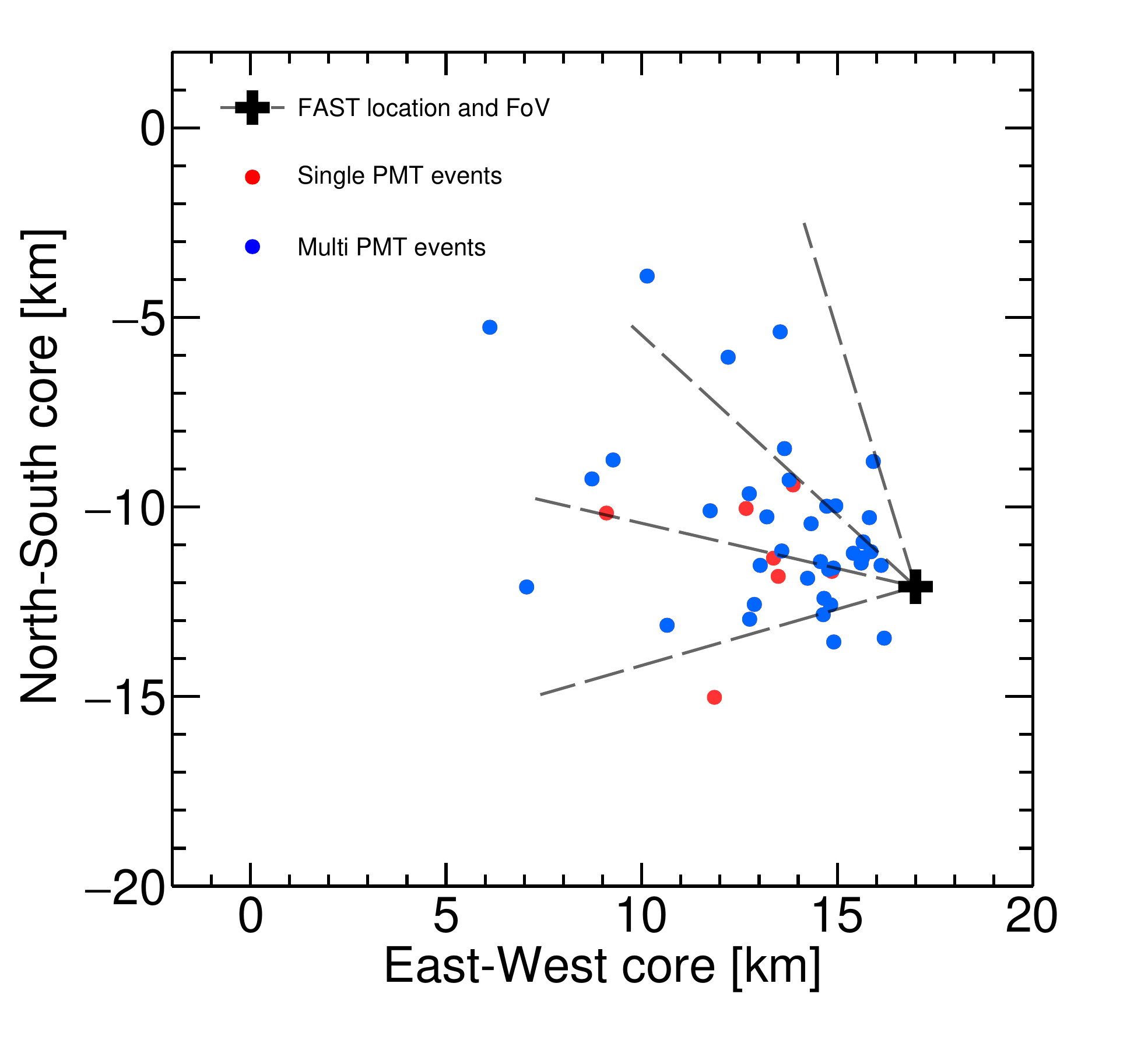}
    \end{subfigure}%
    ~ 
    \begin{subfigure}[t]{0.5\textwidth}
        \centering
        \includegraphics[width=1.\linewidth]{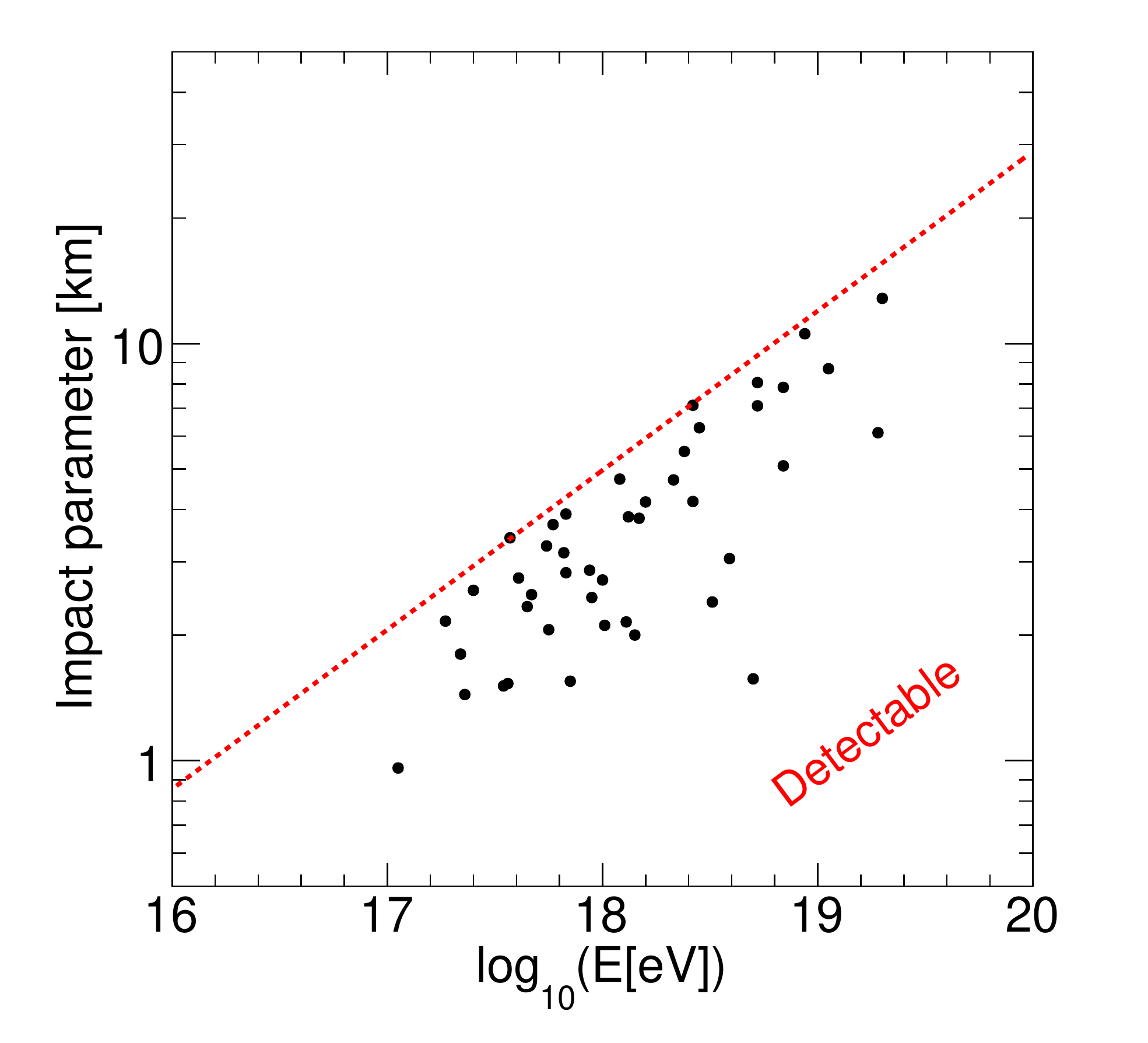}
    \end{subfigure}
    \caption{The 44 events measured in the explored $\sim150$ hour data period. Reconstructed parameters from the Telescope Array monocular reconstruction. Left: Core locations of the events. Right: Distance of closest approach of each shower as a function of reconstructed energy. The red line, fitted to the extreme data points, roughly indicates the maximum detectable shower distance for a given energy.}
    \label{fig:events}
\end{figure}

A total of 44 highly-significant air showers were found in the $\sim150$ hours of explored data. The core locations of the detected showers are shown in the left pane of Fig.~\ref{fig:events}, along with an indication of the field of view of the three telescopes, and the number of pixels in which each shower was detected. While small, this sample provides an estimate of the sensitivity of the full-scale FAST prototypes. The correlation between the distance of closest approach of the shower axis to the FAST prototypes (as determined using the TA monocular reconstruction) and the energy of the 44 showers is plotted in the right pane of Fig.~\ref{fig:events}. We expect showers of a given energy to be detectable up to a maximum impact parameter, which is roughly indicated by the red line. When extrapolated to the FAST target energy of $10^{19.5}$\,eV, a maximum detectable distance of $\sim20$\,km is obtained.

Due to the coarse granularity of the FAST camera necessary for its low cost and straightforward deployment, direct bottom-up reconstruction of measured air showers from the recorded data is not possible; however, an algorithm utilising a top-down approach is currently under development and is discussed in Section~\ref{sec:recon}.


\section{Event simulation and reconstruction}
\label{sec:simRec}


In Sections~\ref{sec:moni} and ~\ref{sec:showers}, we summarised the raw monitoring and scientific measurements taken by the telescope prototypes at the Telescope Array site; here, we summarise the corresponding offline analysis of these data.  As previously discussed, a traditional bottom-up reconstruction that fits for the shower profile and geometry is not possible with the limited timing information provided by FAST; however, a top-down reconstruction method that fits against a library of simulated templates is possible.  A vital ingredient in this process is the development of reliable and quick detector simulation that adequately accounts for the air-shower physics, the various efficiencies of the optical apparatus, and the ray-tracing simulation of the telescope.

In this section, we will first discuss the FAST simulation package \verb|FASTSim|, which takes as an input the shower parameters (i.e. energy, geometry, and $X_{\text{max}}$) and returns a full simulation of the measured FAST traces.  We then discuss progress towards the top-down reconstruction procedure, which uses the simulation as a vital ingredient.  Finally, we discuss a technique for fitting atmospheric properties using the simulation, which allows an independent measurement of the extinction properties of the atmosphere using FAST and the CLF.

\subsection{Simulation of the FAST prototype telescope}
\label{sec:sim}

A full end-to-end simulation of the FAST prototype telescopes is implemented in a modified version of the Pierre Auger Observatory's \Offline software framework~\cite{bib:augeroffline}. The \Offline framework is written in C++, and allows for the straightforward implementation of simulation algorithms via self-contained physics- and detector-related \textit{modules}. The event simulation is driven by an XML steering file, known as a \textit{module sequence}, which provides the list of modules to be run and the order in which to run them. Upon completion, each module relays the status of the event to the next module in the sequence. The run parameters of each module are set in dedicated steering files. 


The energy deposited in the atmosphere by the shower as a function of slant depth is based on a simple Gaisser-Hillas parameterisation~\cite{bib:gh_function} and is dependent on the chosen shower energy, depth of maximum ($X_{\text{max}}$), zenith angle, and two additional parameters controlling the shape of the profile. 
The shower core is placed on the ground with respect to the FAST prototype telescope. A coordinate system consistent with that of the Telescope Array experiment is currently implemented, facilitating straightforward comparison between simulated events and real events measured at the TA site. The number of fluorescence photons at the shower track is calculated using the AIRFLY fluorescence model~\cite{bib:airfly1,bib:airfly2}, with height-dependent air temperature, pressure, and humidity profiles provided by a realistic parameterisation of a typical desert atmosphere. The Cherenkov photon contribution is derived from the number of shower electrons above the Cherenkov threshold in air, calculated from the energy deposit profile using the inverse of the mean ionisation loss rate~\cite{bib:nerling2005}.

An end-to-end detector simulation is performed in the \verb|FASTSimulator| module. Photons (direct fluorescence, Rayleigh-scattered Cherenkov, Mie-scattered Cherenkov, and direct Cherenkov) are propagated in a wavelength-dependent way through a parameterised molecular and aerosol atmosphere to each FAST telescope. The simulation subsequently propagates light through the FAST optics using a combination of simulated and measured properties of its spectral response. This includes accounting for the UV filter transmission and mirror reflectivity, as well as the telescope's directional sensitivity and optical spot size based on the full raytracing simulation presented in Section~\ref{sec:perf}. The signal in each PMT as a function of time is calculated in units of photoelectrons per 100\,ns using the laboratory-measured azimuthally-dependent detection efficiency of the PMTs. The location, pointing direction, and number of FAST telescopes can be altered via the module's steering card.

\begin{figure}[t]
    \centering
    \begin{subfigure}{0.5\textwidth}
        \centering
        \includegraphics[width=1.\linewidth]{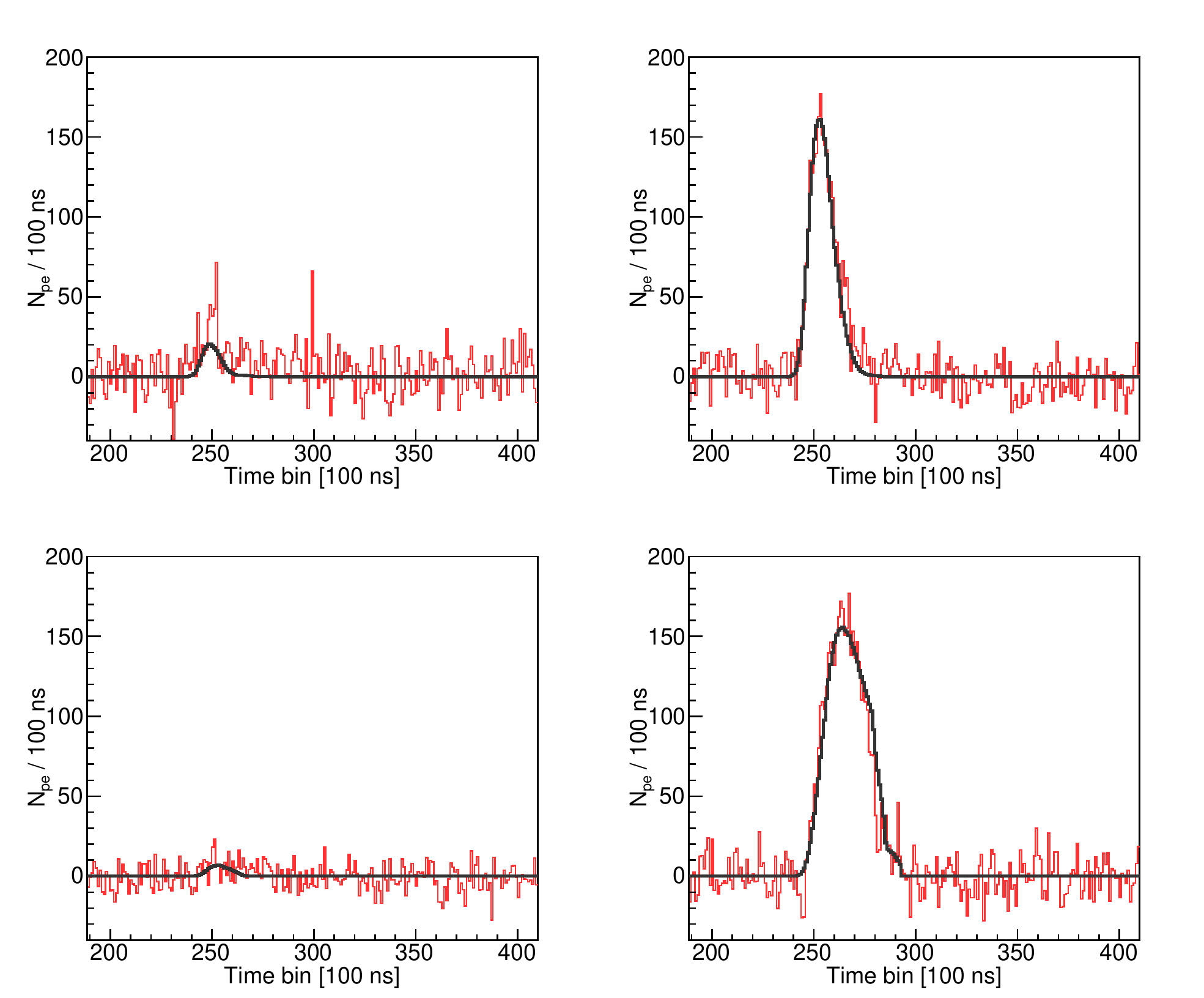}
    \end{subfigure}%
    ~ 
    \begin{subfigure}{0.5\textwidth}
        \centering
        \includegraphics[width=1.\linewidth]{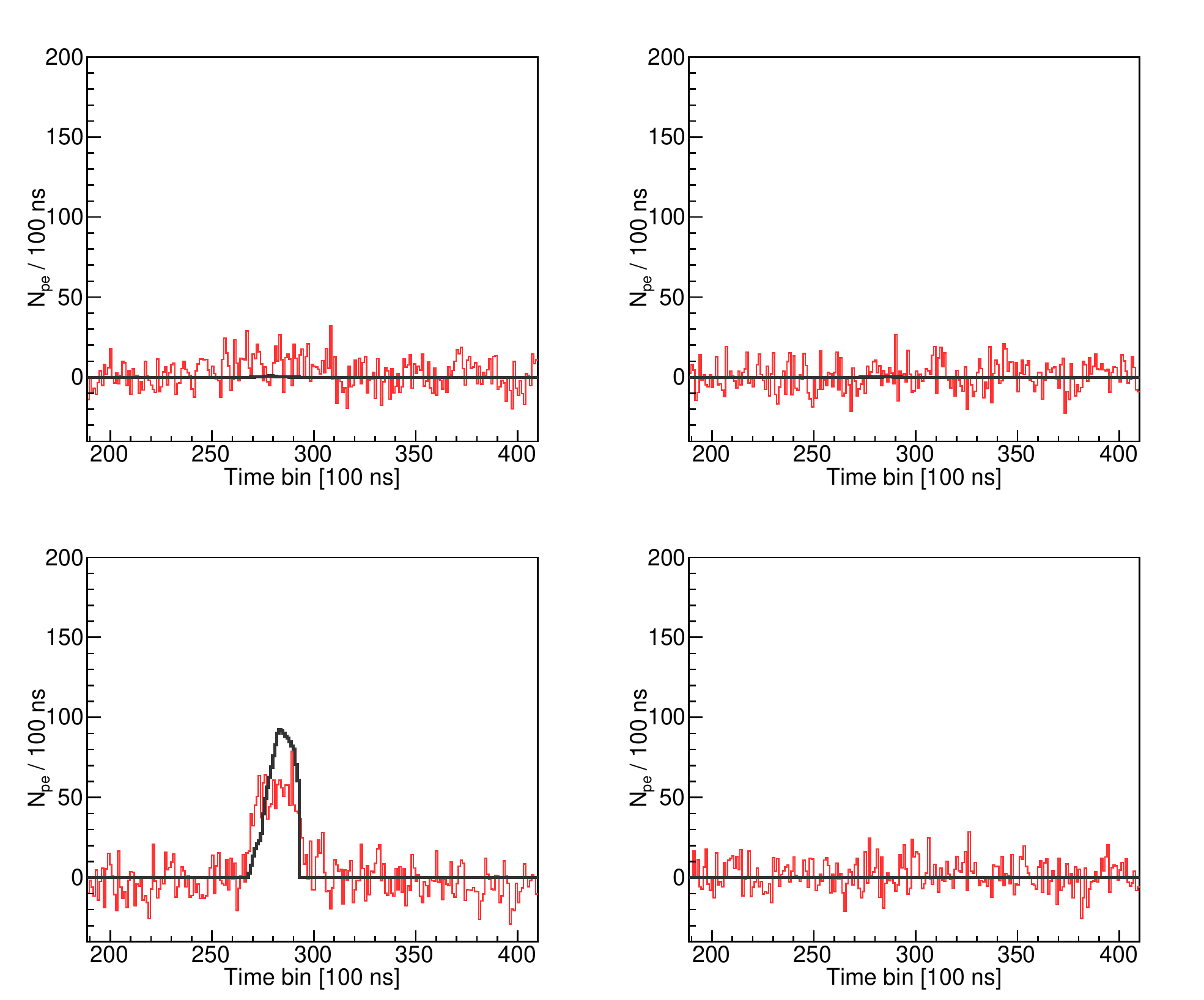}
    \end{subfigure}
    \caption{A simulation of the measured event depicted in Fig~\ref{fig:shower} based on the best-fit parameters given by the top-down event reconstruction in black, overlaid on the measured FAST event in red.}
    \label{fig:simex}
\end{figure}

An example of a simulation of the measured FAST event discussed in Section~\ref{sec:showers} is shown in Fig.~\ref{fig:simex}. The measured event is shown in red, while the simulated prediction for the best-fit shower parameters supplied by the top-down reconstruction are shown in black. The reconstructed energy and $X_{\text{max}}$ are 17\,EeV and 843\,g/cm$^2$, respectively. Good agreement can be seen between the shape and amplitude of the simulated and measured data. The offset in the normalisation of the signal in one of the lower PMTs is likely due to uncertainties of the calibration on the optics and PMT gains, as this shower has a large direct Cherenkov light contribution due to it being directed towards the telescope.

\subsection{Top-down air shower reconstruction}
\label{sec:recon}

Traditional air shower reconstruction techniques use a bottom-up approach in which only a subset of the available recorded information, such as the total measured signal and the centroid time (signal-weighted time average) of each pixel in the telescope camera, is used to fit the shower parameters. A bottom-up reconstruction typically requires two steps: first, a fit to the shower geometry is performed using the timing information from a track of triggered pixels, and then using this reconstructed shower geometry, the measured light flux is ``unfolded" in order to determine the energy deposited at the shower track, usually expressed in terms of atmospheric slant depth as a Gaisser-Hillas profile. Such a reconstruction is not possible for data recorded with a FAST telescope, since only four pixels cover the same field-of-view as several hundred pixels in a traditional FD telescope. A robust top-down reconstruction algorithm is currently in development which will utilize a sound understanding of the detector response to provide estimates of measured shower parameters with acceptable resolutions.

The top-down approach uses simulations based on first-guess estimates of the shower parameters to perform a maximum-likelihood estimation of the measured shower geometry, energy, and depth of maximum ($X_{\text{max}}$). The maximum-likelihood estimator is built from the probability of measuring a signal of $x_{i}$ photoelectrons in the $i^{\text{th}}$ time bin of FAST pixel $k$, summed over all time bins in the traces of all FAST pixels (including those that did not measure a significant signal). The log-likelihood is given by

\begin{equation}
	\ln \mathcal{L}(\vec{x} \:\vert\: \vec{a}) = \sum_{k}^{N_{\text{pix}}}\sum_{i}^{N_{\text{bins}}} P_{k}(x_i \:\vert\: \vec{a})
\end{equation}

where $\vec{a}$ represents the geometrical and physical parameters ($\theta, \phi, x, y, z, X_{\text{max}}$) of the simulated shower under test. The probability density function for a single pixel with trace length $N_{\text{bins}}$ becomes

\begin{equation}
	P_k(x \:\vert\: \mu_k(\vec{a}), \sigma_{k}, V_{\text{g}}^k) = \sum_{i}^{N_{\text{bins}}} \frac{1}{\sqrt{2\pi [\sigma_{k}^{2} + A\mu_{i}^k(1+V_{\text{g}}^k)]}} \exp \bigg{(} \frac{-(x_i - A\mu_i^k)^2}{2[\sigma_k^2 + A\mu_i^k(1+V_{\text{g}}^k)]} \bigg{)}
\end{equation}

where the expectation value for the observed number of photo-electrons in time bin $i$ of PMT $k$ is given by $\mu_i^k$, and the fluctuations (for large $\mu$) are well-represented by a Gaussian of width $\sqrt{\sigma_k^2 + \mu_i^k(1+V_{\text{g}}^k)}$ where $\sigma_k$ is the baseline variance of the $k^{\text{th}}$ PMT due to the NSB, and $V_{\text{g}}^k$ is the PMT's gain variance. The expected signal is modified by an energy scale factor $A$, a free parameter in control of the energy fit. As the total shower energy simply scales the expected signal, simulating many values of the shower energy is not required.

Preliminary tests suggest geometrical reconstruction utilising this top-down approach will be possible with FAST operating in stereo mode (more than one FAST telescope measuring a single event), while the shower geometry may be provided by a coincident surface detector array for the reconstruction of data from a single FAST telescope. The FAST reconstruction performance and expected resolution are currently being studied using simulated events. 
Due to the computational expense of performing many simulations during this reconstruction procedure, a sound first guess of the shower parameters is necessary in order to minimise the total number of required simulations. A combination of pre-simulated template events and machine learning techniques are being investigated as inputs to a first-guess algorithm.

%

\subsection{Reconstruction of the aerosol loading}
\label{sec:atmosanalysis}

A FAST telescope can be used as an atmospheric monitoring tool by observing ultra-violet laser shots from a distant vertically-fired laser facility. The aerosol content of the atmosphere can be inferred from these measured laser traces through comparisons with simulations and is typically expressed in terms of the vertical aerosol optical depth (VAOD), the integral of the aerosol extinction coefficient $\alpha_{\text{A}}$ from the ground to height $h$

\begin{equation}
	\text{VAOD}(h) = \int_0^h \alpha_{\text{A}}(h') \text{d}h'.
\end{equation}


We have developed a simulation that takes into account the wavelength-dependent attenuation of the 4.4\,mJ, 355\,nm CLF laser beam as it traverses a parameterised atmosphere, as well as the attenuation of light scattered out of the beam towards the FAST telescope aperture. Both the molecular and aerosol atmosphere follow simple exponential models, with the volume scattering coefficient $\alpha$ in each case being described in terms of a ground- or sea-level horizontal attenuation length $L$, and a scale height $H$

\begin{eqnarray}
	\alpha_{\text{M}}(h) &=& (1/L_{\text{M}})\cdot\exp(-(h+H_{\text{g}})/H_{\text{M}}) \nonumber \\
	\alpha_{\text{A}}(h) &=& \begin{cases}
		1/L_{\text{A}} & h < H_{\text{mix}} \\
      (1/L_{\text{A}}) \cdot \exp(-(h-H_{\text{mix}})/H_{\text{A}}) & h \geq H_{\text{mix}}
	\end{cases}
	\label{eqn:atmosmodel}
\end{eqnarray}

The aerosol atmosphere contains an additional mixing layer parameter $H_{\text{mix}}$, allowing for a planetary boundary layer of uniform aerosol density. The molecular horizontal attenuation length at sea-level for a wavelength of 355\,nm is taken from Bucholtz by linearly interpolating between model-determined coefficients at 350\,nm and 360\,nm~\cite{bib:bucholtz}, leading to a sea-level Rayleigh attenuation length of $\sim14.2$\,km. The fraction of laser light scattered towards a FAST telescope from a given height has a dependence on the shape of both the molecular and aerosol scattering phase functions. The shape of the aerosol scattering phase function can not be determined analytically and depends on the size and shape distributions of the aerosols present in the atmosphere. We use the modified Henyey-Greenstein phase function with backscattering parameter $f=0.4$ and asymmetry parameter $g=0.6$, suitable for a dry desert atmosphere, to describe the fraction of laser light per unit solid angle scattered in a particular direction be aerosols~\cite{bib:benzvi2007}.

Following the calculation of the laser light flux at the telescope aperture, the expected signal in photoelectrons is calculated taking into account the measured optical properties of the FAST telescope and the laboratory-measured azimuthally-dependent PMT response. The time-dependent shape and normalisation of the resultant signal encodes information about the attenuation properties of the atmosphere. 

\begin{figure}[htb]
	\centering
	\includegraphics[width=0.7\linewidth]{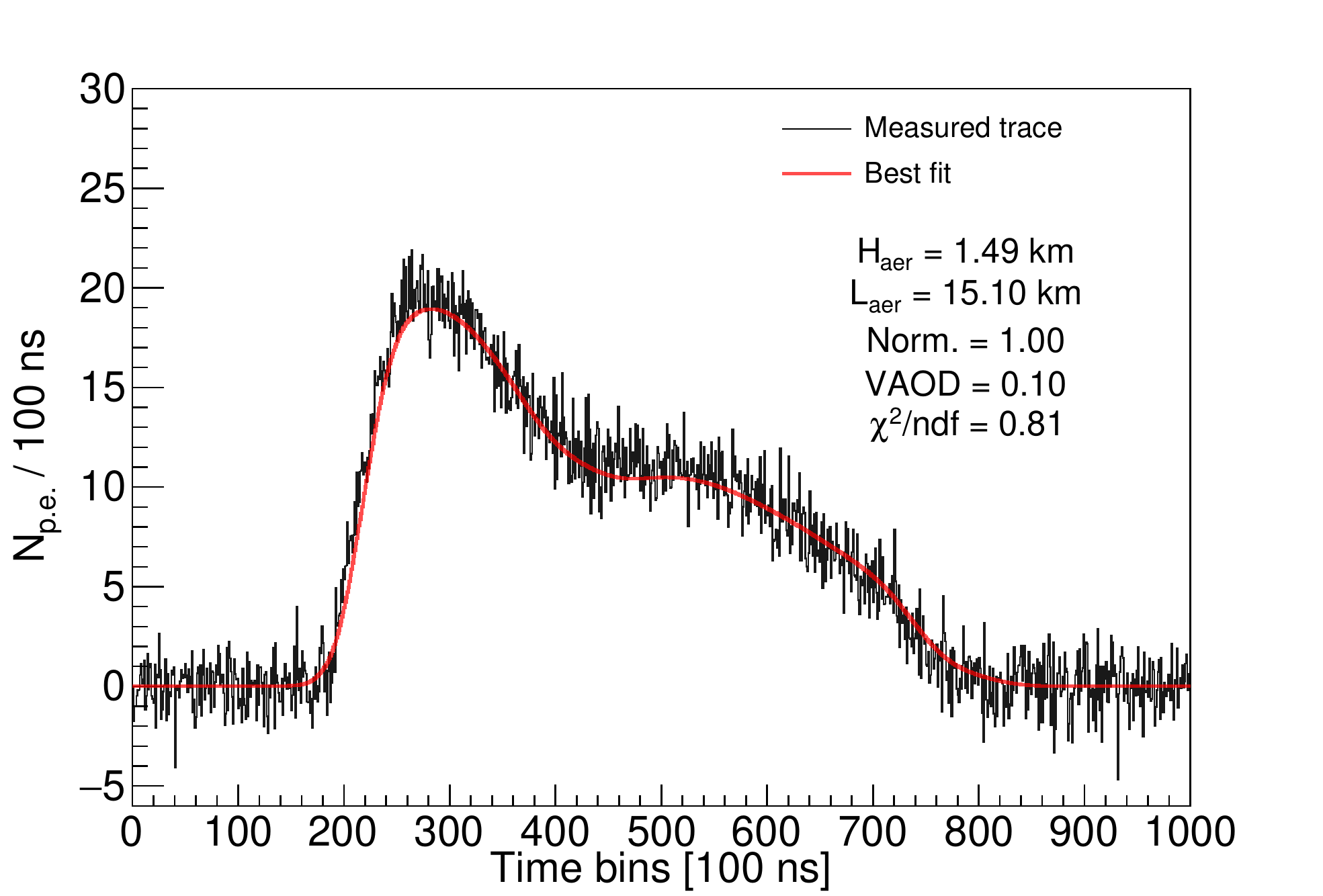}
	\caption{Example reconstruction of the vertical aerosol loading using a simulated FAST laser trace.}
	\label{fig:atmosSim}
\end{figure}

This simulation can be used in conjunction with measurements of the vertically-fired CLF laser at TA to fit for the VAOD at the TA site, which changes over short time-scales due to wind, rain, and other transient atmospheric phenomena. An example application of this VAOD reconstruction procedure is shown in Fig.~\ref{fig:atmosSim}, where a simulation of the expected measured signal due to the TA CLF passing through an aerosol atmosphere with a horizontal attenuation length of 15\,km and a scale height of 1.5\,km is shown in black. The smooth red curve is the result of a simple $\chi^2$ fit to the simulated trace, where the aerosol horizontal attenuation length and scale height are taken as free parameters. This preliminary example shows the potential power of a FAST telescope as an atmospheric monitoring tool.


\section{Future prospects}
\label{sec:future}

\subsection{FAST prototype at the Pierre Auger Observatory}
FAST's low cost, ease of assembly, and autonomous operation lend naturally to the prospect of installing FAST telescopes at multiple sites. This allows one to study environments that may differ in elevation, weather, and atmospheric quality with the same sensor. A natural application of this is cross-calibration of the Telescope Array and Pierre Auger Observatory. For instance, FAST could provide important clues in elucidating the nature of the marked differences in the energy spectrum measured at the Pierre Auger Observatory and the Telescope Array experiment~\cite{bib:spectrumwg2018}, as it is unclear if this can be attributed entirely to differences in the source population (e.g. the TA hotspot) or to a difference in calibrations between the two experiments.  

Though this report primarily deals with the three prototype telescopes installed at TA, in early 2019 we installed a single FAST telescope at the Los Leones site of the Pierre Auger Observatory, which is currently operating (using an internal trigger) in conjunction with the Auger FD.  As discussed in Section~\ref{sec:atmosanalysis}, this allows us to monitor and compare the atmospheres of the two experiments, as well as study showers on an event-by-event basis to compare energy and $X_{\text{max}}$ scales.  FAST also provides the potential to lower the energy threshold of the Los Leones FD by providing a discrete sensor that can detect the Cherenkov light from lower-energy, highly-inclined events.

\subsection{Towards an independent FAST station}
Although the current FAST design is dependent on nearby infrastructure to supply power and network connectivity, the goal of FAST is ultimately a wholly-independent station. Future designs will include a custom-built, self-contained, and solar-powered electronics module capable of providing both the DAQ system and high-voltage for the telescope.  This is currently under development, and we expect that the goal of the next FAST prototype will be testing of this design.

Continued operation of the existing detectors is also a vital step towards achieving a full FAST array, because it is critical that we understand the evolution and degradation of the detectors as they age.  We must know that a FAST telescope can achieve its science goals while operating independently, without maintenance, for years at a time.  To this end, we will continue to run the existing FAST prototypes in coincidence with the TA and Auger FD.

%

\section{Conclusion}
\label{sec:conclusion}

Over the past few years, we have demonstrated the feasibility and reliability of the FAST model of fluorescence detection, with the ultimate goal of laying the foundation for a future array with an order of magnitude larger ground coverage than previous-generation detectors targeted at the highest-energy cosmic rays. We have measured UHECRs with energies above $10^{19}$\,eV and have analysed vertical laser signals to investigate the atmospheric transparency above the detector. Further, we have shown a novel method for event reconstruction that allows us to circumvent one of the principal limitations of a coarsely-pixelised camera: the lack of timing information to tightly constrain the shower geometry. Continued operation will allow us to further test the robustness of a FAST telescope while we work towards achieving full independence from existing FD infrastructure, and in the process, FAST telescopes installed at both the Telescope Array and Pierre Auger Observatory sites will allow us to compare the quality of the atmosphere and sky between the two largest current-generation detectors.

\section*{Acknowledgments}
This work was supported by JSPS KAKENHI Grant Number 18KK0381, 18H01225, 15H05443,
and Grant-in-Aid for JSPS Research Fellow 16J04564 and JSPS Fellowships H25-339, H28-4564, and a research grant from the Hakubi Center for Advanced Research, Kyoto University.
This work was partially carried out by the joint research program of
the Institute for Cosmic Ray Research (ICRR) at the University of Tokyo.
This work was supported in part by NSF grant PHY-1713764, PHY-1412261 and by the Kavli
Institute for Cosmological Physics at the University of Chicago through
grant NSF PHY-1125897 and an endowment from the Kavli Foundation and its founder Fred Kavli.
The Czech authors gratefully acknowledge the support of the Ministry of Education,
Youth and Sports of the Czech Republic project No. LTAUSA17078, CZ.02.1.01/0.0/17\_049/0008422, LTT 18004. The Australian authors acknowledge the support of the Australian Research Council, through Discovery Project DP150101622.
The authors thank the Telescope Array and Pierre Auger Collaborations for productive discussions, and for providing logistical support and part of the instrumentation required to perform the FAST prototype measurements in the field. 

\bibliography{fast2019}

\begin{thebibliography}{10}
\expandafter\ifx\csname url\endcsname\relax
  \def\url#1{\texttt{#1}}\fi
\expandafter\ifx\csname urlprefix\endcsname\relax\def\urlprefix{URL }\fi
\expandafter\ifx\csname href\endcsname\relax
  \def\href#1#2{#2} \def\path#1{#1}\fi

\bibitem{bib:TA}
M.~Fukushima, et~al., {Telescope array project for extremely high energy cosmic
  rays}, Prog.Theor.Phys.Suppl. 151 (2003) 206--210.
\newblock \href {http://dx.doi.org/10.1143/PTPS.151.206}
  {\path{doi:10.1143/PTPS.151.206}}.

\bibitem{bib:ThePierreAuger:2015rma}
A.~Aab, et~al., {The Pierre Auger Cosmic Ray Observatory}, Nucl. Instrum. Meth.
  A798 (2015) 172--213.
\newblock \href {http://arxiv.org/abs/1502.01323} {\path{arXiv:1502.01323}},
  \href {http://dx.doi.org/10.1016/j.nima.2015.06.058}
  {\path{doi:10.1016/j.nima.2015.06.058}}.

\bibitem{bib:augerSD}
I.~Allekotte, et~al., {The Surface Detector System of the Pierre Auger
  Observatory}, Nucl.Instrum.Meth. A586 (2008) 409--420.
\newblock \href {http://arxiv.org/abs/0712.2832} {\path{arXiv:0712.2832}},
  \href {http://dx.doi.org/10.1016/j.nima.2007.12.016}
  {\path{doi:10.1016/j.nima.2007.12.016}}.

\bibitem{bib:tasd}
T.~Abu-Zayyad, et~al., {The surface detector array of the Telescope Array
  experiment}, Nucl.Instrum.Meth. A689 (2012) 87--97.
\newblock \href {http://arxiv.org/abs/1201.4964} {\path{arXiv:1201.4964}},
  \href {http://dx.doi.org/10.1016/j.nima.2012.05.079}
  {\path{doi:10.1016/j.nima.2012.05.079}}.

\bibitem{bib:augerFD}
J.~Abraham, et~al., {The Fluorescence Detector of the Pierre Auger
  Observatory}, Nucl.Instrum.Meth. A620 (2010) 227--251.
\newblock \href {http://arxiv.org/abs/0907.4282} {\path{arXiv:0907.4282}},
  \href {http://dx.doi.org/10.1016/j.nima.2010.04.023}
  {\path{doi:10.1016/j.nima.2010.04.023}}.

\bibitem{bib:tafd}
H.~Tokuno, Y.~Tameda, M.~Takeda, K.~Kadota, D.~Ikeda, et~al., {New air
  fluorescence detectors employed in the Telescope Array experiment},
  Nucl.Instrum.Meth. A676 (2012) 54--65.
\newblock \href {http://arxiv.org/abs/1201.0002} {\path{arXiv:1201.0002}},
  \href {http://dx.doi.org/10.1016/j.nima.2012.02.044}
  {\path{doi:10.1016/j.nima.2012.02.044}}.

\bibitem{bib:gzk1}
K.~Greisen, {End to the cosmic ray spectrum?}, Phys.Rev.Lett. 16 (1966)
  748--750.
\newblock \href {http://dx.doi.org/10.1103/PhysRevLett.16.748}
  {\path{doi:10.1103/PhysRevLett.16.748}}.

\bibitem{bib:gzk2}
G.~Zatsepin, V.~Kuzmin, {Upper limit of the spectrum of cosmic rays}, JETP
  Lett. 4 (1966) 78--80.

\bibitem{Aab:2017cgk}
A.~Aab, et~al., {Inferences on mass composition and tests of hadronic
  interactions from 0.3 to 100 EeV using the water-Cherenkov detectors of the
  Pierre Auger Observatory}, Phys. Rev. D96~(12) (2017) 122003.
\newblock \href {http://arxiv.org/abs/1710.07249} {\path{arXiv:1710.07249}},
  \href {http://dx.doi.org/10.1103/PhysRevD.96.122003}
  {\path{doi:10.1103/PhysRevD.96.122003}}.

\bibitem{bib:tahotspot}
R.~U. Abbasi, et~al., {Indications of Intermediate-Scale Anisotropy of Cosmic
  Rays with Energy Greater Than 57 EeV in the Northern Sky Measured with the
  Surface Detector of the Telescope Array Experiment}, Astrophys. J. 790 (2014)
  L21.
\newblock \href {http://arxiv.org/abs/1404.5890} {\path{arXiv:1404.5890}},
  \href {http://dx.doi.org/10.1088/2041-8205/790/2/L21}
  {\path{doi:10.1088/2041-8205/790/2/L21}}.

\bibitem{bib:aniso_wg2018}
J.~Biteau, et~al., {Covering the celestial sphere at ultra-high energies}, EPJ
  Web Conf. 210 (2019) 01005.
\newblock \href {http://arxiv.org/abs/1905.04188} {\path{arXiv:1905.04188}}.

\bibitem{bib:augerdipole}
A.~Aab, et~al., {Observation of a Large-scale Anisotropy in the Arrival
  Directions of Cosmic Rays above $8 \times 10^{18}$ eV}, Science 357~(6537)
  (2017) 1266--1270.
\newblock \href {http://arxiv.org/abs/1709.07321} {\path{arXiv:1709.07321}},
  \href {http://dx.doi.org/10.1126/science.aan4338}
  {\path{doi:10.1126/science.aan4338}}.

\bibitem{bib:augerdipole2}
A.~Aab, et~al., {Large-scale cosmic-ray anisotropies above 4 EeV measured by
  the Pierre Auger Observatory}, Astrophys. J. 868~(1) (2018) 4.
\newblock \href {http://arxiv.org/abs/1808.03579} {\path{arXiv:1808.03579}},
  \href {http://dx.doi.org/10.3847/1538-4357/aae689}
  {\path{doi:10.3847/1538-4357/aae689}}.

\bibitem{bib:firstfast}
T.~Fujii, et~al., {Detection of ultra-high energy cosmic ray showers with a
  single-pixel fluorescence telescope}, Astropart. Phys. 74 (2016) 64--72.
\newblock \href {http://arxiv.org/abs/1504.00692} {\path{arXiv:1504.00692}},
  \href {http://dx.doi.org/10.1016/j.astropartphys.2015.10.006}
  {\path{doi:10.1016/j.astropartphys.2015.10.006}}.

\bibitem{bib:fastoptics}
D.~Mandat, M.~Palatka, M.~Pech, P.~Schovanek, P.~Travnicek, L.~Nozka,
  M.~Hrabovsky, P.~Horvath, T.~Fujii, P.~Privitera, M.~Malacari, J.~Farmer,
  A.~Galimova, A.~Matalon, M.~Merolle, X.~Ni, J.~Bellido, J.~Matthews,
  S.~Thomas, \href{http://stacks.iop.org/1748-0221/12/i=07/a=T07001}{The
  prototype opto-mechanical system for the fluorescence detector array of
  single-pixel telescopes}, Journal of Instrumentation 12~(07) (2017) T07001.
\newline\urlprefix\url{http://stacks.iop.org/1748-0221/12/i=07/a=T07001}

\bibitem{bib:chibacalib}
R.~Abbasi, et~al., {Calibration and Characterization of the IceCube
  Photomultiplier Tube}, Nucl. Instrum. Meth. A618 (2010) 139--152.
\newblock \href {http://arxiv.org/abs/1002.2442} {\path{arXiv:1002.2442}},
  \href {http://dx.doi.org/10.1016/j.nima.2010.03.102}
  {\path{doi:10.1016/j.nima.2010.03.102}}.

\bibitem{bib:TAtrigger}
Y.~Tameda, et~al., {Trigger electronics of the new fluorescence detectors of
  the Telescope Array experiment}, Nucl. Instrum. Meth. A609 (2009) 227--234.
\newblock \href {http://dx.doi.org/10.1016/j.nima.2009.07.093}
  {\path{doi:10.1016/j.nima.2009.07.093}}.

\bibitem{bib:gemmeke2003}
H.~Gemmeke, M.~Kleifges, A.~Menshikov, {Statistical calibration and background
  measurements of the Auger fluorescence detector} (2003) 891--894.

\bibitem{Shin:2014uha}
B.~K. Shin, et~al., {Gain monitoring of telescope array photomultiplier cameras
  for the first 4 years of operation}, Nucl. Instrum. Meth. A768 (2014)
  96--103.
\newblock \href {http://dx.doi.org/10.1016/j.nima.2014.09.059}
  {\path{doi:10.1016/j.nima.2014.09.059}}.

\bibitem{bib:EScale}
V.~Verzi, {The Energy Scale of the Pierre Auger Observatory}, in: {Proceedings,
  33rd International Cosmic Ray Conference (ICRC2013): Rio de Janeiro, Brazil,
  July 2-9, 2013}, p. 0928.

\bibitem{bib:airfly2}
M.~Ave, et~al., {Measurement of the pressure dependence of air fluorescence
  emission induced by electrons}, Astropart.Phys. 28 (2007) 41--57.
\newblock \href {http://arxiv.org/abs/astro-ph/0703132}
  {\path{arXiv:astro-ph/0703132}}, \href
  {http://dx.doi.org/10.1016/j.astropartphys.2007.04.006}
  {\path{doi:10.1016/j.astropartphys.2007.04.006}}.

\bibitem{bib:clf}
S.~Udo, M.~Allen, R.~Cady, M.~Fukushima, Y.~Iida, J.~N.~Matthews, et~al., {The
  Central Laser Facility at the Telescope Array}, Proc. of the 30th
  International Cosmic Ray Conference, Merida, Mexico 5 (2007) 1021--1024.

\bibitem{bib:ctaasc}
D.~Mandat, et~al., {All Sky Camera for the CTA Atmospheric Calibration work
  package}, EPJ Web Conf. 89 (2015) 03007.
\newblock \href {http://dx.doi.org/10.1051/epjconf/20158903007}
  {\path{doi:10.1051/epjconf/20158903007}}.

\bibitem{bib:augeroffline}
S.~Argiro, S.~L.~C. Barroso, J.~Gonzalez, L.~Nellen, T.~C. Paul, T.~A. Porter,
  L.~Prado, Jr., M.~Roth, R.~Ulrich, D.~Veberic, {The Offline Software
  Framework of the Pierre Auger Observatory}, Nucl. Instrum. Meth. A580 (2007)
  1485--1496.
\newblock \href {http://arxiv.org/abs/0707.1652} {\path{arXiv:0707.1652}},
  \href {http://dx.doi.org/10.1016/j.nima.2007.07.010}
  {\path{doi:10.1016/j.nima.2007.07.010}}.

\bibitem{bib:gh_function}
T.~K.~Gaisser, A.~M.~Hillas, {Reliability of the method of constant intensity
  cuts for reconstructing the average development of vertical showers},
  International Cosmic Ray Conference, 15th ICRC (Plovdiv) 8 (1977) 353--357.

\bibitem{bib:airfly1}
M.~Ave, et~al., {Precise measurement of the absolute fluorescence yield of the
  337 nm band in atmospheric gases}, Astropart.Phys. 42 (2013) 90--102.
\newblock \href {http://arxiv.org/abs/1210.6734} {\path{arXiv:1210.6734}},
  \href {http://dx.doi.org/10.1016/j.astropartphys.2012.12.006}
  {\path{doi:10.1016/j.astropartphys.2012.12.006}}.

\bibitem{bib:nerling2005}
F.~Nerling, J.~Bluemer, R.~Engel, M.~Risse, {Universality of electron
  distributions in high-energy air showers: Description of Cherenkov light
  production}, Astropart. Phys. 24 (2006) 421--437.
\newblock \href {http://arxiv.org/abs/astro-ph/0506729}
  {\path{arXiv:astro-ph/0506729}}, \href
  {http://dx.doi.org/10.1016/j.astropartphys.2005.09.002}
  {\path{doi:10.1016/j.astropartphys.2005.09.002}}.

\bibitem{bib:bucholtz}
A.~{Bucholtz}, {Rayleigh-scattering calculations for the terrestrial
  atmosphere}, Applied Optics 34~(15) (1995) 2765--2773.

\bibitem{bib:benzvi2007}
S.~Y. {BenZvi}, B.~M. {Connolly}, J.~A.~J. {Matthews}, M.~{Prouza}, E.~F.
  {Visbal}, S.~{Westerhoff}, {Measurement of the aerosol phase function at the
  Pierre Auger Observatory}, Astroparticle Physics 28~(3) (2007) 312--320.

\bibitem{bib:spectrumwg2018}
T.~AbuZayyad, et~al., {Auger-TA energy spectrum working group report}, EPJ Web
  Conf. 210 (2019) 01002.
\newblock \href {http://dx.doi.org/10.1051/epjconf/201921001002}
  {\path{doi:10.1051/epjconf/201921001002}}.

\end{thebibliography}

\end{document}